\documentclass[
reprint,
superscriptaddress,
amsmath,amssymb,
aps,
pra,
10pt
]{revtex4-2}

\usepackage{graphicx}
\usepackage{xcolor}
\usepackage{tikz-cd}
\usepackage{etoolbox}
\makeatletter
\patchcmd{\@bibdataout@aps}{author="08"}{author="48"}{}{} % chktex 18
\makeatother

\begin{document}

\title{Breakdown of the Overdamped Approximation in Fluctuating Environments}

\author{Kazuki Fukutani}
\affiliation{Department of Physics and Astronomy, Tokyo University of Science,
Noda, Chiba 278-8510, Japan}

\author{Takuma Akimoto}
\email{takuma@rs.tus.ac.jp}
\affiliation{Department of Physics and Astronomy, Tokyo University of Science,
Noda, Chiba 278-8510, Japan}

\date{\today}

\begin{abstract}
The overdamped approximation is widely used to describe Brownian
motion in complex environments, but its validity becomes nontrivial
when the friction coefficient itself fluctuates in time.
We investigate this problem by comparing underdamped and overdamped
Langevin dynamics subject to the same fluctuating friction and
satisfying the fluctuation--dissipation relation at a common
temperature.
We show that environmental averaging and inertial elimination
generally lead to different long-time transport when environmental
fluctuations are fast compared with velocity relaxation.
For rapidly fluctuating friction, the underdamped dynamics is governed
by the arithmetic mean friction and yields
$D_{\rm eff}^{\rm under}=k_{\rm B}T/\langle\gamma\rangle$,
whereas the overdamped dynamics gives
$D_{\rm eff}^{\rm over}=k_{\rm B}T\langle\gamma^{-1}\rangle$.
For a two-state Markov friction, we derive the finite-time effective
diffusion coefficient exactly and identify the full crossover between
these two regimes, controlled by the competition between velocity
relaxation and environmental switching.
We further establish the fast-fluctuation result for general
stationary friction processes and verify it for a continuous
log-Ornstein--Uhlenbeck environment, for which the characteristic
crossover time can also be determined independently.
Our results reveal a noncommutativity between environmental averaging
and inertial elimination, establish a timescale-dependent criterion
for the validity of overdamped dynamics in temporally heterogeneous
environments, and show that rapid environmental fluctuations can
enhance, rather than suppress, the consequences of inertia.
\end{abstract}

\maketitle

\section{Introduction}

%Overdamped approximation
The overdamped approximation is one of the most widely used reductions
in stochastic dynamics \cite{kramers1940brownian,risken1989fokker}.
For a Brownian particle with a constant friction coefficient, inertia
becomes irrelevant on timescales longer than the velocity-relaxation
time, leading to the familiar overdamped Langevin equation.
High-bandwidth and high-resolution measurements have provided direct
experimental access to the inertial regime of Brownian motion, resolving
ballistic motion, the crossover to diffusion, and hydrodynamic memory
effects
\cite{huang2011direct,franosch2011resonances,brites2016instantaneous}.
In active systems, inertia can have even more persistent consequences:
the coupling of velocity relaxation to additional dynamical degrees of
freedom can affect transport beyond the short-time ballistic regime
\cite{scholz2018inertial,lowen2020inertial,nguyen2022active}.
The elimination of inertia becomes more subtle, however, when the
friction or noise amplitude depends on the state of the system.
In particular, studies of the Smoluchowski--Kramers limit have shown
that a naive removal of the velocity degree of freedom can miss
noise-induced drift terms in systems with nonuniform friction and
diffusion
\cite{sancho1982adiabatic,hottovy2012noise,hottovy2015smoluchowski}.
These results demonstrate that eliminating inertia in heterogeneous
environments can require more care than simply neglecting the
acceleration term.

%Dynamical heterogeneity in complex systems
A distinct issue arises when the mobility experienced by a diffusing
particle itself evolves in time. Temporal mobility fluctuations arise
in a wide variety of systems, either because the surrounding environment
changes dynamically or because the diffusing object undergoes changes
in its internal state, conformation, or shape
\cite{wang2009anomalous,wang2012brownian,Jeon2012,He2013,
Bhattacharya2013,Guan2014,kwon2014dynamics,Manzo2015,Doi-Edwards-book,Yamamoto2021}.
In supercooled and glass-forming liquids, crowded fluids, and other
complex media, for example, a probe particle may successively experience
local environments with markedly different mobilities
\cite{kwon2014dynamics,Yamamoto-Onuki-1998,Yamamoto-Onuki-1998a,
Kob1997,kawasaki2007correlation,Berthier2011,hachiya2019unveiling,
Miotto2021,sposini2023glassy}.
In either case, the mobility relevant to particle motion evolves on
a characteristic timescale, introducing an additional dynamical
timescale alongside the intrinsic relaxation of the particle.

%Fluctuating-diffusivity models
Temporal mobility fluctuations have motivated stochastic descriptions
in which the diffusivity itself evolves in time, including
fluctuating-diffusivity, diffusing-diffusivity, and related models
\cite{Chubynsky2014,Uneyama2015,Miyaguchi2016,Chechkin2017,Massignan2014, AkimotoYamamoto2016a,kimura2023non,shirataki2026observation,Akimoto2026}. Most such descriptions are formulated at the overdamped 
level and have been successful in describing non-Gaussian displacement
statistics and other signatures of heterogeneous transport \cite{Chubynsky2014,Uneyama2015,Miyaguchi2016,Chechkin2017,Massignan2014, AkimotoYamamoto2016a,Akimoto2026,Akimoto2016,Miyaguchi2019,Uneyama2019}.
Underdamped dynamics with temporally fluctuating friction has also been
considered in several contexts
\cite{rozenfeld1998brownian,luczka2000diffusion,Perakis2026}.
These studies demonstrate that coupling inertia to a fluctuating
environment can produce nontrivial dynamical behavior. However, the
conditions under which such underdamped dynamics can be consistently
reduced to an overdamped fluctuating-mobility description, and whether
the two descriptions yield the same long-time transport coefficient,
remain much less understood.
Interpreting such a description as an
overdamped reduction of an underlying inertial dynamics requires the
velocity to relax sufficiently rapidly compared with changes in the
local mobility. When the environmental and velocity-relaxation
timescales become comparable, this separation of timescales breaks down:
the mobility may change before the velocity has equilibrated locally,
allowing inertial memory to persist across environmental changes.
Whether eliminating the velocity degree of freedom can then alter
transport even in the long-time diffusive regime remains an open
question.

%conceptual problem
Recent mathematical studies of multiscale Langevin equations have shown
that homogenization of a rapidly fluctuating environment and the
Smoluchowski--Kramers limit need not commute \cite{hua2026smoluchowski,qian2026smoluchowski}.
This raises a complementary physical question: at finite mass and finite
environmental relaxation time, can the competition between these
timescales produce observable differences between underdamped and
overdamped transport, and when does the overdamped approximation become
quantitatively inaccurate?
In the underdamped dynamics, the velocity responds over a finite
relaxation time to the fluctuating friction, whereas eliminating inertia
first makes the particle respond instantaneously to the local mobility.
Because mobility is the inverse of friction, these two descriptions
perform fundamentally different averaging operations when environmental
and velocity-relaxation timescales compete.
It therefore remains to determine how this competition controls the
long-time transport coefficient and whether rapid environmental
fluctuations suppress or preserve the effects of inertia.

%Our results
We show that slow environmental fluctuations allow the velocity to
equilibrate locally and recover the overdamped result, whereas rapid
fluctuations produce a persistent discrepancy even in the long-time
diffusive regime. In the fast-fluctuation limit, the underdamped
dynamics is governed by the arithmetic mean friction,
$D_{\rm eff}^{\rm under}=k_{\rm B}T/\langle\gamma\rangle$, whereas
eliminating inertia first gives
$D_{\rm eff}^{\rm over}=k_{\rm B}T\langle\gamma^{-1}\rangle$, where
$\gamma$ denotes the fluctuating friction coefficient.
The resulting difference reflects the noncommutativity of environmental
averaging and inertial elimination. For a two-state Markov environment,
we derive the full finite-time crossover exactly and obtain a
timescale criterion for the validity of the overdamped approximation.
We then show that the fast-fluctuation result extends to general
fluctuating-friction processes, demonstrating that the breakdown is
not specific to dichotomous switching.

\section{Setup}
\label{sec: model}

We consider a one-dimensional underdamped Brownian particle subject to a
temporally fluctuating friction coefficient $\gamma(t)$. Its dynamics is
described by
\begin{align}
  \dot{x}(t) &= v(t), \label{eq:kinematic}\\
  m\dot{v}(t)
  &=
  -\gamma(t)v(t)
  +\sqrt{2k_{\rm B}T\gamma(t)}\,\xi(t),
  \label{eq:ULE}
\end{align}
where $m$ is the particle mass, $T$ is the temperature, and $\xi(t)$ is
Gaussian white noise satisfying
\begin{equation}
  \langle\xi(t)\rangle=0,
  \qquad
  \langle\xi(t)\xi(t')\rangle=\delta(t-t').
\end{equation}
The friction coefficient $\gamma(t)>0$ represents temporal fluctuations
of the local environment and is assumed to be statistically independent
of the thermal noise $\xi(t)$. The noise amplitude in
Eq.~\eqref{eq:ULE} satisfies the fluctuation--dissipation relation
instantaneously at the same temperature $T$ for every realization of
$\gamma(t)$.

The corresponding overdamped description is obtained by eliminating the
velocity degree of freedom while retaining the same fluctuating
environment,
\begin{equation}
  \dot{x}(t)
  =
  \sqrt{\frac{2k_{\rm B}T}{\gamma(t)}}\,\xi(t).
  \label{eq:OD_model}
\end{equation}
Thus, the instantaneous local diffusivity in the overdamped description is
\begin{equation}
  D(t)=\frac{k_{\rm B}T}{\gamma(t)}.
  \label{eq:local_diffusivity}
\end{equation}
Equations~\eqref{eq:ULE} and~\eqref{eq:OD_model} provide the common
framework for comparing the underdamped and overdamped descriptions.
Importantly, the same stochastic friction process $\gamma(t)$ is used in
both models, so that any difference between them arises solely from the
elimination of the velocity degree of freedom.

For both descriptions, we characterize transport over a finite observation
time $t$ by
\begin{equation}
  D_{\rm eff}(t)
  \equiv
  \frac{\left\langle [x(t)-x(0)]^2\right\rangle}{2t}.
  \label{eq:Deff_finite_def}
\end{equation}
We denote the corresponding quantities by
$D_{\rm eff}^{\rm under}(t)$ and $D_{\rm eff}^{\rm over}(t)$. 

We set the initial position to $x(0)=0$ and draw the initial velocity
from the Maxwell--Boltzmann distribution at temperature $T$,
\begin{equation}
  \langle v(0)\rangle=0,
  \qquad
  \langle v^2(0)\rangle=\frac{k_{\rm B}T}{m}.
\end{equation}
The initial velocity is assumed to be statistically independent of the
initial friction coefficient $\gamma(0)$.

 \if0
The following two properties of the model will be used repeatedly
in Sec.~III:
\begin{enumerate}
  \item[(A1)] The joint process $(v(t),\sigma(t))$ is Markovian.
  The environmental state switches with constant rates $k_l$ and $k_h$,
  independently of the velocity, and $v(t)$ remains continuous at
  switching events.

  \item[(A2)] Both environmental states satisfy the
  fluctuation--dissipation relation at the same temperature $T$.
  The initial velocity is Maxwellian and statistically independent
  of $\sigma(0)$.
\end{enumerate}

For an arbitrary state-dependent quantity $a_w$, we define
\begin{equation}
  \langle a\rangle\equiv\sum_{w=l,h}P_w a_w,
  \qquad
  \langle a\rangle_t\equiv\sum_{w=l,h}p_w(t)a_w.
  \label{eq:environmental_averages}
\end{equation}
Equation~\eqref{eq:p_relax_active} gives
\begin{equation}
  \langle a\rangle_t
  =\langle a\rangle+\Delta q\,(a_l-a_h)e^{-Kt}.
  \label{eq:environmental_average_relax}
\end{equation}
Equation~\eqref{eq:environmental_average_relax} shows that any
state-dependent quantity relaxes exponentially toward its stationary
average with the environmental relaxation rate $K$. The initial
preparation affects only the amplitude of this transient contribution,
through $\Delta q(a_l-a_h)$, while its relaxation time $K^{-1}$ is
independent of the quantity considered. 
\fi

\section{Exact results for two-state friction}
\label{sec:two_state}

We first specialize the fluctuating friction to a two-state Markov
process. The environmental state is represented by
$\sigma(t)\in\{l,h\}$, and the friction coefficient is
\begin{equation}
  \gamma(t)=\gamma_{\sigma(t)}
  =
  \begin{cases}
    \gamma_l, & \sigma(t)=l,\\
    \gamma_h, & \sigma(t)=h,
  \end{cases}
  \qquad \gamma_l<\gamma_h .
  \label{eq:gamma_switch}
\end{equation}
At a switching event, only the friction coefficient changes, with no
impulsive force acting on the particle; hence the velocity remains
continuous across the switch.

For a fixed environmental state $w\in\{l,h\}$, the velocity dynamics
reduces to an Ornstein--Uhlenbeck process with velocity-relaxation time
and diffusion coefficient
\begin{equation}
  \tau_{v,w}\equiv\frac{m}{\gamma_w},
  \quad
  D_w\equiv\frac{k_{\rm B}T}{\gamma_w}.
  \label{eq:tauv_D_def}
\end{equation}
Thus, $\gamma_l<\gamma_h$ implies
$\tau_{v,l}>\tau_{v,h}$ and $D_l>D_h$.

We assume exponentially distributed residence times in each
environmental state,
\begin{equation}
  P(\Delta t_w)
  =
  \frac{1}{\tau_w}
  \exp\!\left(-\frac{\Delta t_w}{\tau_w}\right),
  \label{eq:waiting_time}
\end{equation}
where $\tau_w$ is the mean residence time and
$k_w\equiv\tau_w^{-1}$ the corresponding escape rate.
The transitions $l\to h$ and $h\to l$ therefore occur with rates
$k_l$ and $k_h$, respectively. The stationary occupation
probabilities are
\begin{equation}
  P_l=\frac{k_h}{k_l+k_h}
      =\frac{\tau_l}{\tau_l+\tau_h},
  \quad
  P_h=\frac{k_l}{k_l+k_h}
      =\frac{\tau_h}{\tau_l+\tau_h}.
  \label{eq:stationary_prob}
\end{equation}
The environmental relaxation rate is
\begin{equation}
  K\equiv k_l+k_h
  =\frac{1}{\tau_l}+\frac{1}{\tau_h},
  \label{eq:K_def}
\end{equation}
so that $K^{-1}$ is the environmental relaxation time.

We denote the occupation probability of environmental state $w$ by
\begin{equation}
  p_w(t)\equiv \Pr[\sigma(t)=w].
  \label{eq:occupation_prob}
\end{equation}
The initial environmental state is specified by $p_l(0)$ and
$p_h(0)=1-p_l(0)$.
The occupation probabilities then relax toward their
stationary values as
\begin{equation}
  p_l(t)=P_l+\Delta q\,e^{-Kt},
  \quad
  p_h(t)=P_h-\Delta q\,e^{-Kt},
  \label{eq:p_relax_active}
\end{equation}
where
\begin{equation}
  \Delta q\equiv p_l(0)-P_l
  =-\bigl[p_h(0)-P_h\bigr].
  \label{eq:Deltaq_def}
\end{equation}
Here $-P_l\leq\Delta q\leq P_h$, and $\Delta q=0$ corresponds to
stationary initialization.

For the two-state model, we make the dependence on the environmental
relaxation rate explicit and write the finite-time diffusion coefficients
as $D_{\rm eff}^{\rm under}(t,K)$ and
$D_{\rm eff}^{\rm over}(t,K)$. Their long-time limits are denoted by
\begin{equation}
  D_{\rm eff}^{\alpha}(K)
  \equiv
  \lim_{t\to\infty}D_{\rm eff}^{\alpha}(t,K),
  \quad
  \alpha=\mathrm{under},\mathrm{over}.
  \label{eq:Deff_models}
\end{equation}
In the following, we first derive the general finite-time diffusion
coefficients and then examine their behavior in the long-time,
fast-switching, and slow-switching regimes.

\subsection{General finite-time theory}

We first derive the velocity correlation function required to obtain the
finite-time effective diffusion coefficient of the underdamped model.
Although the velocity $v(t)$ alone is generally non-Markovian, the joint
process $(v(t),\sigma(t))$ is Markovian. This property allows us to derive
closed evolution equations for velocity correlations resolved with respect
to the environmental state.

For a reference time $t_1$ and a lag time $s\ge0$, we define the
velocity correlation function as
\begin{equation}
  C(t_1,t_1+s)  \equiv  \langle v(t_1)v(t_1+s)\rangle .
  \label{eq:C_def}
\end{equation}
We further decompose the velocity correlation according to the
environmental state at the later time and define, for $w\in\{l,h\}$, the state-resolved
velocity correlations
\begin{equation}
  C_w(t_1,t_1+s)=\bigl\langle v(t_1)v(t_1+s)\,
  \mathbf{1}_{\{\sigma(t_1+s)=w\}}\bigr\rangle ,
  \label{eq:Cw_def}
\end{equation}
where $\mathbf{1}_{\{\cdot\}}$ denotes the indicator function. The total
velocity correlation is then  $C(t_1,t_1+s)=C_l(t_1,t_1+s)+C_h(t_1,t_1+s)$.
Since the joint process $(v(t),\sigma(t))$ is Markovian, the
state-resolved correlations obey a closed system of evolution
equations with respect to the lag time $s$:
\begin{align}
  \frac{\partial C_l}{\partial s}
  &=
  -\left(\frac{\gamma_l}{m}+k_l\right)C_l
  +k_h C_h,
  \label{eq:dCf}\\
  \frac{\partial C_h}{\partial s}
  &=
  k_l C_l
  -\left(\frac{\gamma_h}{m}+k_h\right)C_h.
  \label{eq:dCs}
\end{align}
Here and below, the arguments $(t_1,t_1+s)$ of $C_w$ are omitted
when no confusion arises. A derivation based on a short-time path
decomposition is given in Appendix~\ref{app: Expansion}.

Equations~\eqref{eq:dCf} and \eqref{eq:dCs} can be written in matrix form as
\begin{equation}
  \frac{\partial}{\partial s}
  \begin{pmatrix}
    C_l\\ C_h
  \end{pmatrix}
  =
  \mathsf{M}
  \begin{pmatrix}
    C_l\\ C_h
  \end{pmatrix},
  \qquad
  \mathsf{M}
  =
  \begin{pmatrix}
    -a & k_h\\
    k_l & -b
  \end{pmatrix},
  \label{eq:matrix_form}
\end{equation}
where
\begin{equation}
  a=\frac{1}{\tau_{v,l}}+k_l,
  \qquad
  b=\frac{1}{\tau_{v,h}}+k_h.
\end{equation}
The eigenvalues of $\mathsf{M}$ are
\begin{equation}
  \lambda_\pm
  =
  -\frac{a+b}{2}
  \pm
  \frac{1}{2}
  \sqrt{(a-b)^2+4k_lk_h}.
  \label{eq:eigenvalues}
\end{equation}
The two eigenvalues are negative and determine the characteristic
decay rates of the velocity correlation.

\if0
Using $K$, $P_l$ and $P_h$ defined in Eqs.~\eqref{eq:stationary_prob}
and~\eqref{eq:K_def}, they can be rewritten as
\begin{align}
  \lambda_\pm
  ={}&-\frac{1}{2}\left(K+\frac{1}{\tau_{v,l}}+\frac{1}{\tau_{v,h}}\right)
  \notag\\
  &\pm\frac{1}{2}\sqrt{K^{2}+2(P_h-P_l)K\,\Delta K_v+\Delta K_v^{2}},
  \label{eq:eigenvalues_PK}
\end{align}
where
\begin{equation}
  \Delta K_v=\frac{1}{\tau_{v,l}}-\frac{1}{\tau_{v,h}}
  \label{eq:DeltaKv}
\end{equation}
is the difference between the velocity-relaxation rates in the two states.

Since $k_lk_h>0$, the discriminant in Eq.~\eqref{eq:eigenvalues_PK} is a sum
of squares,
$\bigl[K+(P_h-P_l)\Delta K_v\bigr]^2+4P_lP_h\Delta K_v^2>0$
(using $(P_h-P_l)^2+4P_lP_h=1$), so $\lambda_\pm$ are real and distinct for
all parameter values and the bi-exponential form \eqref{eq:C_general} below
holds without degenerate special cases.
\fi

The total velocity correlation therefore takes the form
\begin{equation}
  C(t_1,t_1+s)
  =
  A_+(t_1)e^{\lambda_+s}
  +
  A_-(t_1)e^{\lambda_-s},
  \label{eq:C_general}
\end{equation}
where the coefficients $A_\pm(t_1)$ are determined by the state-resolved
second moments of the velocity, 
\begin{equation}
  \Phi_w(t)
  \equiv
  \left\langle
  v^2(t)\mathbf{1}_{\{\sigma(t)=w\}}
  \right\rangle .
  \label{eq:Phi_def}
\end{equation}
By definition, the equal-time state-resolved correlations satisfy
$C_w(t,t)=\Phi_w(t)$.
Since the initial velocity is drawn from the Maxwell--Boltzmann
distribution independently of the environmental state,
\begin{equation}
  \Phi_w(0)  =  \frac{k_{\rm B}T}{m}p_w(0) .
  \label{eq:init_Phi}
\end{equation}
The equal-time state-resolved correlations can also be obtained exactly.
Their evolution equations are
\begin{align}
  \frac{d\Phi_l}{dt}
  &=
  -\left(\frac{2\gamma_l}{m}+k_l\right)\Phi_l
  +k_h\Phi_h
  +\frac{2k_{\rm B}T\gamma_l}{m^2}p_l(t),
  \label{eq:dPhil}\\
  \frac{d\Phi_h}{dt}
  &=
  k_l\Phi_l
  -\left(\frac{2\gamma_h}{m}+k_h\right)\Phi_h
  +\frac{2k_{\rm B}T\gamma_h}{m^2}p_h(t).
  \label{eq:dPhih}
\end{align}
A derivation is given in Appendix~\ref{app: Phi}.
For the Maxwell--Boltzmann initial velocity distribution introduced in
Sec.~\ref{sec: model}, these equations have the simple solution
\begin{equation}
  \Phi_w(t)
  =
  \frac{k_{\rm B}T}{m}p_w(t).
  \label{eq:Phi_equipartition}
\end{equation}
Thus, even when the environmental occupation probabilities are
nonstationary, the conditional velocity variance remains at the
equipartition value,
\begin{equation}
  \left\langle v^2(t)\mid\sigma(t)=w\right\rangle
  =
  \frac{k_{\rm B}T}{m}.
  \label{eq:conditional_equipartition}
\end{equation}
Consequently, $\langle v^2(t)\rangle=k_{\rm B}T/m$ at all times.
From Eq.~\eqref{eq:Phi_equipartition}, the equal-time velocity
correlation is
\begin{equation}
  C(t_1,t_1)=\frac{k_{\rm B}T}{m}.
  \label{eq:C_ic}
\end{equation}
Summing Eqs.~\eqref{eq:dCf} and \eqref{eq:dCs} and evaluating the
result at $s=0$ yields
\begin{equation}
  \left.
  \frac{\partial C(t_1,t_1+s)}{\partial s}
  \right|_{s=0}
  =
  -\frac{k_{\rm B}T}{m^2}
  \langle\gamma(t_1)\rangle ,
  \label{eq:dC_dtau_ic}
\end{equation}
where
$\langle\gamma(t_1)\rangle
=p_l(t_1)\gamma_l+p_h(t_1)\gamma_h$.
Thus, the equal-time value is fixed by equipartition, whereas the
initial decay rate is determined by the instantaneous mean friction.
 
%\paragraph{Determination of $A_\pm(t_1)$.}
Evaluating Eq.~\eqref{eq:C_general} and its derivative with respect
to $s$ at $s=0$ and using Eqs.~\eqref{eq:C_ic} and
\eqref{eq:dC_dtau_ic}, we obtain
\begin{align}
  A_+(t_1)
  &=
  \frac{k_{\rm B}T}{m}
  \frac{-\langle\gamma(t_1)\rangle/m-\lambda_-}
       {\lambda_+-\lambda_-},
  \\
  A_-(t_1)
  &=
  \frac{k_{\rm B}T}{m}
  \frac{\langle\gamma(t_1)\rangle/m+\lambda_+}
       {\lambda_+-\lambda_-}.
  \label{eq:Apm_general}
\end{align}
For a general initial environmental distribution, the mean friction
relaxes as
\begin{equation}
  \langle\gamma(t_1)\rangle
  =
  \langle\gamma\rangle_{\rm eq}
  +
  \delta\gamma_0 e^{-Kt_1},
  \label{eq:gamma_relax}
\end{equation}
where $\langle\gamma\rangle_{\rm eq}=P_l\gamma_l+P_h\gamma_h$ and 
  $\delta\gamma_0 \equiv
  \langle\gamma(0)\rangle-\langle\gamma\rangle_{\rm eq}$.
Accordingly, the coefficients can be decomposed as
\begin{equation}
  A_\pm(t_1)
  =
  A_\pm^{\rm eq}
  +
  \delta A_\pm e^{-Kt_1},
  \label{eq:Apm}
\end{equation}
with
\begin{align}
  A_+^{\rm eq}
  &=
  \frac{k_{\rm B}T}{m}
  \frac{-\langle\gamma\rangle_{\rm eq}/m-\lambda_-}
       {\lambda_+-\lambda_-},
  \\
  A_-^{\rm eq}
  &=
  \frac{k_{\rm B}T}{m}
  \frac{\langle\gamma\rangle_{\rm eq}/m+\lambda_+}
       {\lambda_+-\lambda_-},
  \label{eq:Apm_eq}
\end{align}
and
\begin{equation}
  \delta A_+
  =
  -\frac{k_{\rm B}T}{m^2}
  \frac{\delta\gamma_0}{\lambda_+-\lambda_-},
  \quad
  \delta A_-=-\delta A_+.
  \label{eq:dApm}
\end{equation}
Thus, the $t_1$ dependence of $A_\pm(t_1)$ arises solely from the
relaxation of the environmental occupation probabilities.

%\paragraph{Mean-squared displacement and finite-time diffusion coefficient.}

The displacement over time $t$ is
$\Delta x(t)=\int_0^t v(t')\,dt'$. Using the symmetry of the velocity
correlation, the MSD can be written as
\begin{equation}
  \left\langle [x(t)-x(0)]^2\right\rangle
  =
  2\int_0^t dt_1
  \int_0^{t-t_1} ds\,
  C(t_1,t_1+s).
  \label{eq:MSD_def}
\end{equation}
The expressions derived above are valid for an arbitrary initial
environmental distribution. 
Substituting Eqs.~\eqref{eq:C_general} and
\eqref{eq:Apm} into Eq.~\eqref{eq:MSD_def} and carrying out
the integrations, we obtain
\begin{equation}
  D_{\rm eff}^{\rm under}(t,K)
  =
  D_{\rm eq}(t,K)
  +
  D_{\rm tr}(t,K),
  \label{eq:Dunder}
\end{equation}
where
\begin{equation}
  D_{\rm eq}(t,K)
  =
  -\sum_{\alpha=\pm}
  \frac{A_\alpha^{\rm eq}}{\lambda_\alpha}
  +
  \frac{1}{t}
  \sum_{\alpha=\pm}
  \frac{A_\alpha^{\rm eq}}
       {\lambda_\alpha^2}
  \left(e^{\lambda_\alpha t}-1\right)
  \label{eq:Deq}
\end{equation}
is the contribution for stationary environmental initialization, while
\begin{equation}
  D_{\rm tr}(t,K)
  =
  \frac{1}{t}
  \sum_{\alpha=\pm}
  \frac{\delta A_\alpha}{\lambda_\alpha}
  \left[
    \frac{e^{\lambda_\alpha t}-e^{-Kt}}
         {\lambda_\alpha+K}
    -
    \frac{1-e^{-Kt}}{K}
  \right]
  \label{eq:Dtr}
\end{equation}
is the transient contribution arising from a nonstationary initial
environmental distribution.
The transient term is proportional to
$\delta\gamma_0=\langle\gamma(0)\rangle-\langle\gamma\rangle_{\rm eq}$
and therefore vanishes for stationary initialization. Moreover,
$D_{\rm tr}(t,K)\to0$ as $t\to\infty$, so that the long-time diffusion
coefficient is independent of the initial environmental preparation.

%\paragraph{Overdamped counterpart.}
For comparison, we now derive the finite-time diffusion coefficient
of the overdamped model defined in Eq.~\eqref{eq:OD_model}.
The ensemble-averaged instantaneous diffusivity is
\begin{equation}
  \langle D(t')\rangle
  =
  \langle D\rangle_{\rm eq}
  +
  \delta D_0 e^{-Kt'},
  \label{eq:Dinst_over}
\end{equation}
where  $\langle D\rangle_{\rm eq}=  P_lD_l+P_hD_h$ and 
$\delta D_0 \equiv \langle D(0)\rangle-\langle D\rangle_{\rm eq}$.
Since
$\left\langle [x(t)-x(0)]^2\right\rangle
=2\int_0^t\langle D(t')\rangle\,dt'$,
the finite-time effective diffusion coefficient is
\begin{equation}
  D_{\rm eff}^{\rm over}(t,K)
  =
  \langle D\rangle_{\rm eq}
  +
  \delta D_0
  \frac{1-e^{-Kt}}{Kt}.
  \label{eq:Dover_general}
\end{equation}
In particular, the long-time overdamped diffusion coefficient is
\begin{equation}
  D_{\rm eff}^{\rm over}
  =
  \langle D\rangle_{\rm eq}
  =
  k_{\rm B}T
  \langle\gamma^{-1}\rangle_{\rm eq},
  \label{eq:Dover_long}
\end{equation}
independent of the switching rate $K$ and of the initial
environmental preparation.
 
\subsection{Long-time breakdown of the overdamped approximation}
\label{sec:Deff}

We first consider the long-time limit at a fixed switching rate $K>0$.
Throughout this subsection, we vary the switching time scale through
$K$ while keeping the stationary occupations $P_l$ and $P_h$ fixed,
i.e., $k_l=KP_h$ and $k_h=KP_l$.
In the limit $t\to\infty$, the transient contribution arising from a
nonstationary initial environmental distribution vanishes. The long-time
diffusion coefficient is therefore independent of the initial environmental
preparation and is given by
\begin{equation}
  D_{\rm eff}^{\rm under}(K)
  =
  -\left(
  \frac{A_+^{\rm eq}}{\lambda_+}
  +
  \frac{A_-^{\rm eq}}{\lambda_-}
  \right).
  \label{eq:Dunder_tinf_eig}
\end{equation}
Using the sum and product of the eigenvalues,
\begin{align}
  \lambda_++\lambda_-
  &=
  -\left(
  \frac{1}{\tau_{v,l}}
  +\frac{1}{\tau_{v,h}}
  +K
  \right),
  \\
  \lambda_+\lambda_-
  &=
  \frac{1}{\tau_{v,l}\tau_{v,h}}
  +K\frac{\langle\gamma\rangle_{\rm eq}}{m},
  \label{eq:eig_sum_prod}
\end{align}
we obtain
\begin{equation}
  D_{\rm eff}^{\rm under}(K)
  =
  \frac{k_{\rm B}T}{m}
  \frac{
  \dfrac{P_h}{\tau_{v,l}}
  +\dfrac{P_l}{\tau_{v,h}}
  +K}
  {
  \dfrac{1}{\tau_{v,l}\tau_{v,h}}
  +K\dfrac{\langle\gamma\rangle_{\rm eq}}{m}
  }.
  \label{eq:Dunder_tinf_tau}
\end{equation}
The competition between velocity relaxation and environmental switching
is characterized by the dimensionless parameter
\begin{equation}
  \Theta
  \equiv
  K\tau_{v,l}\tau_{v,h}
  \frac{\langle\gamma\rangle_{\rm eq}}{m}
  =
  \frac{\tau_{v,l}}{\tau_l}
  +
  \frac{\tau_{v,h}}{\tau_h}.
  \label{eq:Theta_gamma}
\end{equation}
Thus, $\Theta$ compares the velocity-relaxation time in each
environmental state with the corresponding mean residence time.
The limits $\Theta\ll1$ and $\Theta\gg1$ therefore describe slow and
fast environmental switching relative to inertial relaxation,
respectively.
In terms of $\Theta$, Eq.~\eqref{eq:Dunder_tinf_tau} takes the compact
form
\begin{equation}
  D_{\rm eff}^{\rm under}(\Theta)
  =
  k_{\rm B}T
  \frac{
  \langle\gamma^{-1}\rangle_{\rm eq}
  +\Theta/\langle\gamma\rangle_{\rm eq}
  }
  {1+\Theta}.
  \label{eq:Dunder_tinf_closed}
\end{equation}
Equation~\eqref{eq:Dunder_tinf_closed} shows that the long-time
diffusion coefficient retains an explicit dependence on the switching
dynamics through $\Theta$. Thus, even after the environmental
occupations have reached their stationary values, the switching time
scale continues to affect diffusion. This dependence originates from
the competition between environmental switching and the finite
velocity-relaxation times: for slow switching, the velocity relaxes
within each environmental state, whereas for fast switching it
experiences many environmental transitions before relaxing. 
Equation~\eqref{eq:Dunder_tinf_closed} therefore describes a crossover
between two distinct friction averages,
$\langle\gamma^{-1}\rangle_{\rm eq}$ and
$1/\langle\gamma\rangle_{\rm eq}$, controlled solely by $\Theta$.

By contrast, the long-time overdamped diffusion coefficient is
independent of $K$,
\begin{equation}
  D_{\rm eff}^{\rm over}
  =
  k_{\rm B}T
  \langle\gamma^{-1}\rangle_{\rm eq}.
  \label{eq:Dover_tinf}
\end{equation}
Consequently, the two descriptions agree in the slow-switching limit,
\begin{equation}
  \lim_{\Theta\to0}
  D_{\rm eff}^{\rm under}
  =
  D_{\rm eff}^{\rm over},
\end{equation}
whereas in the fast-switching limit,
\begin{equation}
  \lim_{\Theta\to\infty}
  D_{\rm eff}^{\rm under}
  =
  \frac{k_{\rm B}T}{\langle\gamma\rangle_{\rm eq}},
  \quad
  D_{\rm eff}^{\rm over}
  =
  k_{\rm B}T\langle\gamma^{-1}\rangle_{\rm eq}.
  \label{eq:fast_limits}
\end{equation}
By Eq.~\eqref{eq:Theta_gamma}, $\Theta\ll1$ means that the velocity relaxes
within each state before the environment escapes it, so that the particle can
adapt locally to every friction state. In the long-time limit,
$\Theta\ll1$ is therefore a sufficient condition for the overdamped
approximation to be valid.

Subtracting Eq.~\eqref{eq:Dunder_tinf_closed} from
Eq.~\eqref{eq:Dover_tinf} yields
\begin{equation}
  \Delta D_{\rm eff}(\Theta)
  =
  k_{\rm B}T
  \left(
    \langle\gamma^{-1}\rangle_{\rm eq}
    -\frac{1}{\langle\gamma\rangle_{\rm eq}}
  \right)
  \frac{\Theta}{1+\Theta},
  \label{eq:DeltaD_general}
\end{equation}
where
$\Delta D_{\rm eff}
\equiv
D_{\rm eff}^{\rm over}-D_{\rm eff}^{\rm under}$.
The difference increases monotonically from zero in the
slow-switching limit ($\Theta\to0$) to
\begin{equation}
  \Delta D_{\rm eff}(\infty)
  =
  k_{\rm B}T
  \left(
    \langle\gamma^{-1}\rangle_{\rm eq}
    -\frac{1}{\langle\gamma\rangle_{\rm eq}}
  \right)
  \label{eq:DeltaD_fast}
\end{equation}
in the fast-switching limit.
Since
$\langle\gamma^{-1}\rangle_{\rm eq}
\ge 1/\langle\gamma\rangle_{\rm eq}$,
the overdamped approximation always overestimates the long-time
diffusion coefficient, with equality only in the absence of friction
heterogeneity.

%\paragraph{Noncommutativity of the overdamped and fast-switching limits.}
The persistence of a finite gap in the fast-switching limit can be
understood from the noncommutativity of the limits $m\to0$ and
$K\to\infty$.
Since $\Theta\propto Km$, Eq.~\eqref{eq:Dunder_tinf_closed} also reveals
that the overdamped and fast-switching limits do not commute. Specifically,
\begin{equation}
  \lim_{K\to\infty}\lim_{m\to0}
  D_{\rm eff}^{\rm under}(K)
  \neq
  \lim_{m\to0}\lim_{K\to\infty}
  D_{\rm eff}^{\rm under}(K)
  \label{eq:m_K_noncommute}
\end{equation}
for a genuinely fluctuating friction.
Taking the overdamped limit first eliminates the velocity degree of
freedom, whereas taking the fast-switching limit first causes the
finite-inertia velocity to experience the averaged friction
$\langle\gamma\rangle_{\rm eq}$.

\begin{figure*}[t]
  \centering
  \includegraphics[width=0.32\textwidth]{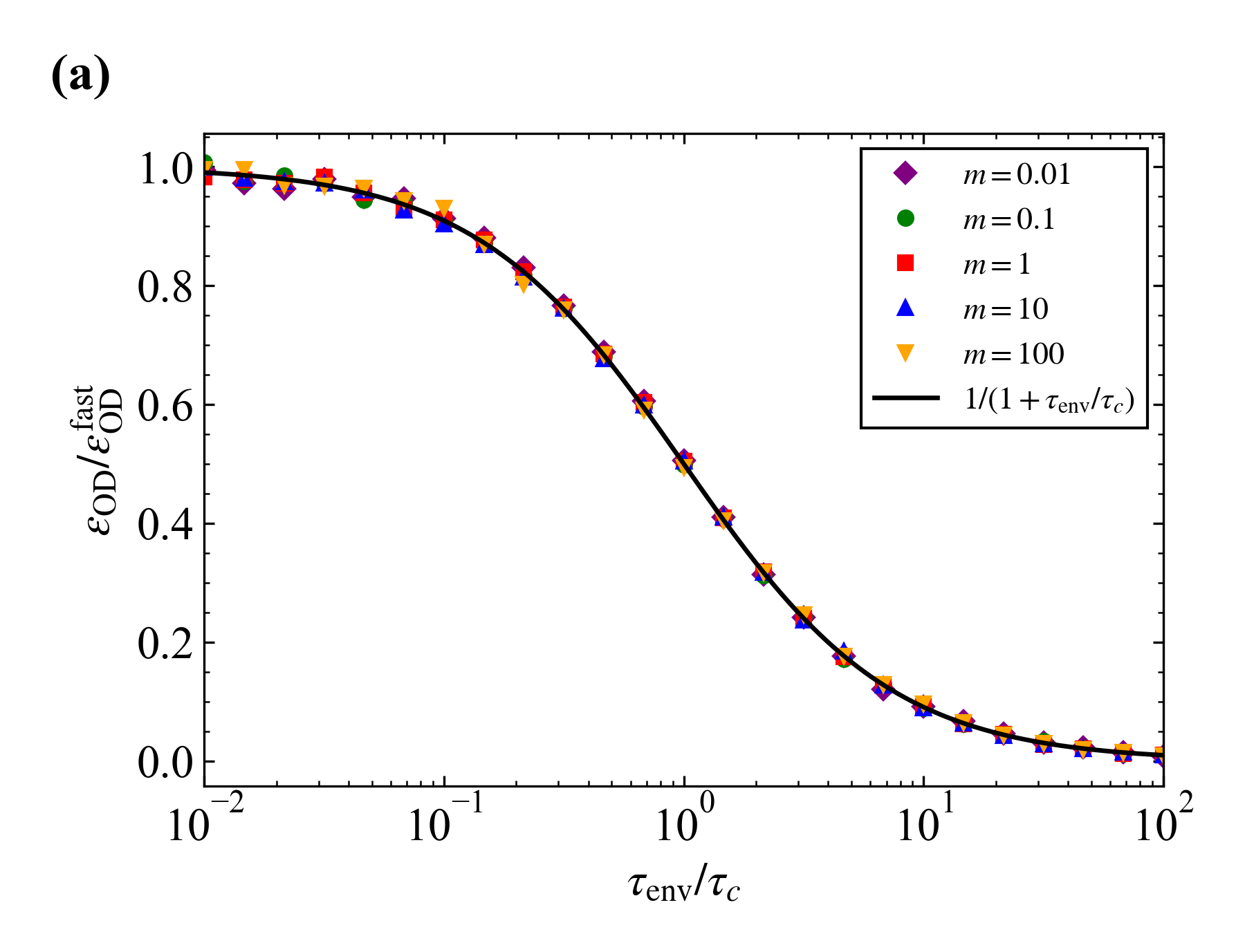}\hfill
  \includegraphics[width=0.32\textwidth]{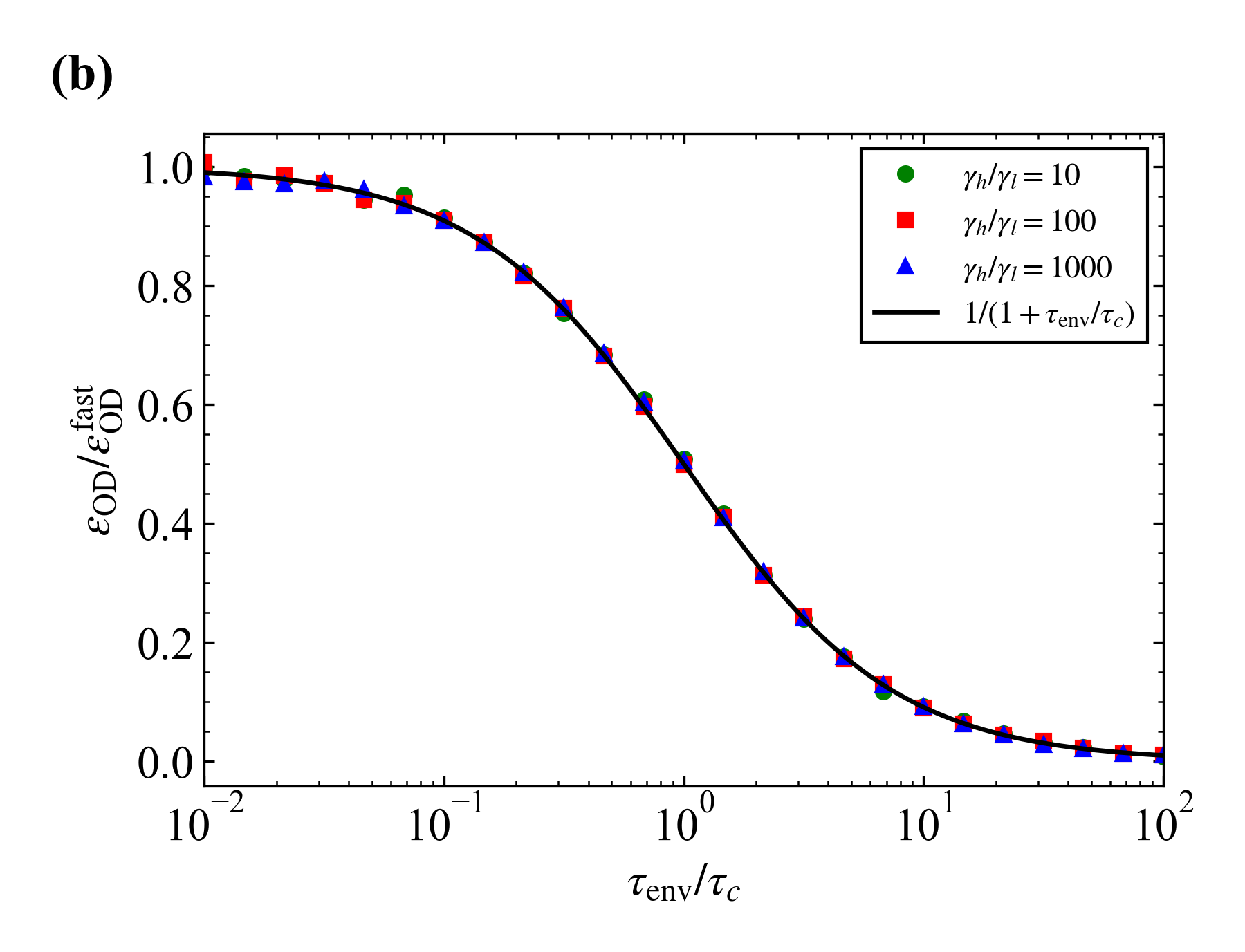}\hfill
  \includegraphics[width=0.32\textwidth]{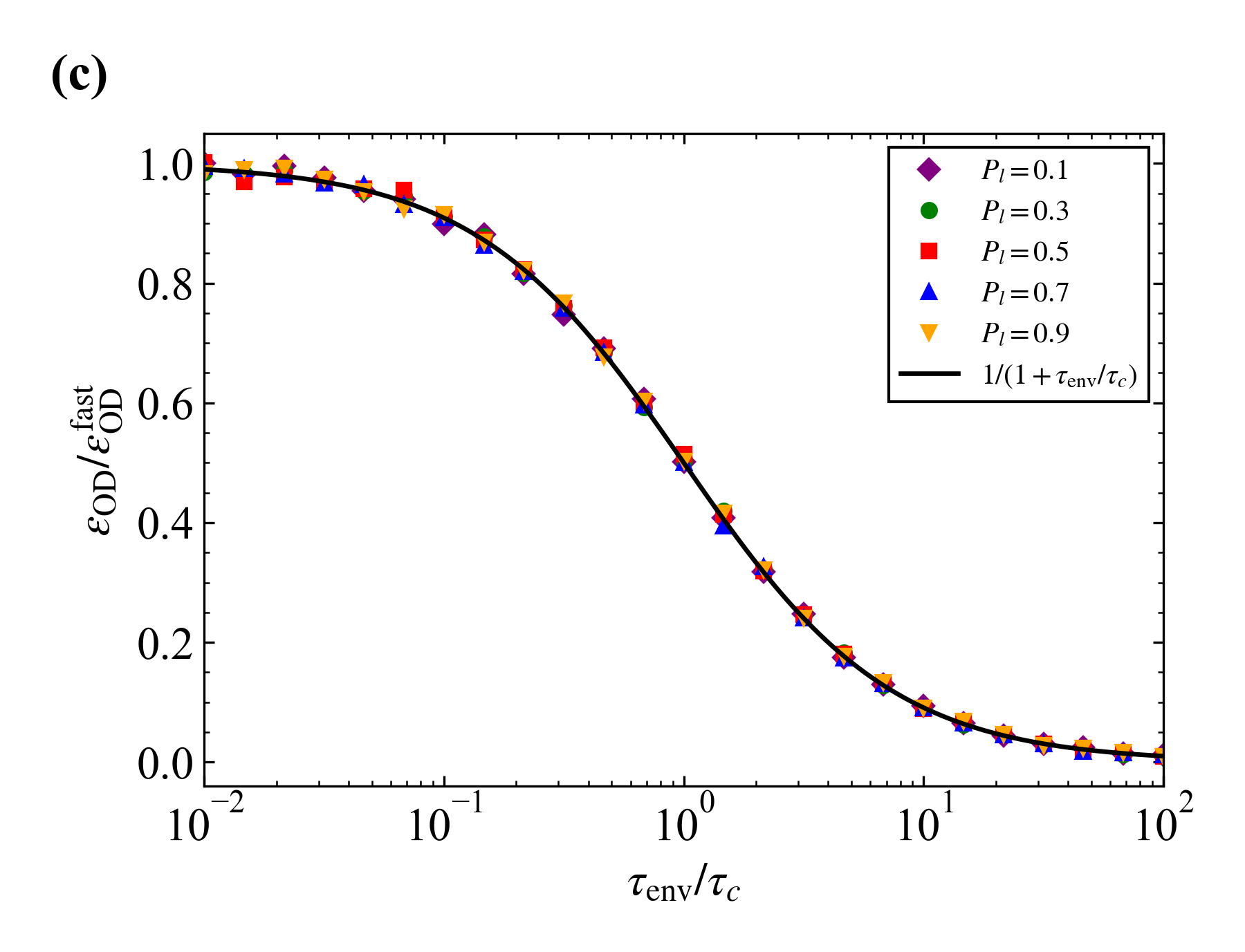}
  \caption{Normalized relative error
$\varepsilon_{\rm OD}/\varepsilon_{\rm OD}^{\rm fast}$ of the overdamped
approximation as a function of the normalized environmental relaxation
time $\tau_{\rm env}/\tau_c$.
(a) Variation with the particle mass $m$.
(b) Variation with the friction ratio $\gamma_h/\gamma_l$.
(c) Variation with the stationary low-friction-state probability $P_l$.
Solid lines show the theoretical prediction
$[1+\tau_{\rm env}/\tau_c]^{-1}$ from Eq.~\eqref{eq:eps_crossover},
while symbols show simulation results.
In all three cases, the data collapse onto the same crossover function.}
  \label{fig:relative_error}
\end{figure*}

To quantify the breakdown of the overdamped approximation, we introduce
the environmental relaxation time
\begin{equation}
  \tau_{\rm env}\equiv K^{-1}
\end{equation}
and define the relative error in the long-time diffusion coefficient as
\begin{equation}
  \varepsilon_{\rm OD}(\tau_{\rm env})
  \equiv
  \frac{D_{\rm eff}^{\rm over}
  -D_{\rm eff}^{\rm under}(\tau_{\rm env})}
  {D_{\rm eff}^{\rm over}}.
  \label{eq:eps_def}
\end{equation}
We further define the crossover time
\begin{equation}
  \tau_c
  \equiv
  P_l\tau_{v,h}+P_h\tau_{v,l}.
  \label{eq:tau_c}
\end{equation}
Since  $\Theta=  K\tau_c=  \tau_c/\tau_{\rm env}$,
Eqs.~\eqref{eq:Dunder_tinf_closed} and \eqref{eq:Dover_tinf} give
\begin{equation}
  \varepsilon_{\rm OD}(\tau_{\rm env})
  =
  \varepsilon_{\rm OD}^{\rm fast}
  \frac{\tau_c}{\tau_{\rm env}+\tau_c}
  ,
  \label{eq:eps_crossover}
\end{equation}
where
\begin{equation}
  \varepsilon_{\rm OD}^{\rm fast}
  =
  1-
  \frac{1}{
  \langle\gamma\rangle_{\rm eq}
  \langle\gamma^{-1}\rangle_{\rm eq}}.
  \label{eq:eps_fast}
\end{equation}
Equation~\eqref{eq:eps_crossover} shows that the crossover is controlled
by the relative magnitude of the environmental relaxation time
$\tau_{\rm env}$ and the characteristic time $\tau_c$. The relative error
reaches half of its fast-switching value at
$\tau_{\rm env}=\tau_c$, vanishes in the slow-switching regime
$\tau_{\rm env}\gg\tau_c$, and saturates at
$\varepsilon_{\rm OD}^{\rm fast}$ in the fast-switching regime
$\tau_{\rm env}\ll\tau_c$.
Figure~\ref{fig:relative_error} demonstrates the resulting scaling
collapse for three distinct parameter variations: the particle mass,
the friction contrast, and the stationary environmental occupation.
Although these variations modify the microscopic inertial and
environmental timescales differently, all data collapse onto the
universal crossover function
$[1+\tau_{\rm env}/\tau_c]^{-1}$.

This crossover provides a simple timescale criterion for the overdamped
approximation. A sufficient condition for its validity is
\begin{equation}
  \tau_{\rm env}\gg\tau_c.
  \label{eq:validity_criterion}
\end{equation}
This condition is sufficient but not necessary for practical accuracy.
When the friction contrast is weak,
$\langle\gamma\rangle_{\rm eq}
\langle\gamma^{-1}\rangle_{\rm eq}\simeq1$, and hence
$\varepsilon_{\rm OD}^{\rm fast}\ll1$; the overdamped approximation can
then remain accurate even when $\tau_{\rm env}\lesssim\tau_c$.
Thus, its accuracy is determined jointly by the competition between
$\tau_{\rm env}$ and $\tau_c$ and by the friction heterogeneity through
$\varepsilon_{\rm OD}^{\rm fast}$.

\subsection{Fast-switching regime}
\label{sec:fast}

We next consider the fast-switching limit $K\to\infty$ at fixed
stationary occupations $P_l$ and $P_h$. The transition rates
$k_l=KP_h$ and $k_h=KP_l$ therefore increase proportionally.
Expanding the eigenvalues in Eq.~\eqref{eq:eigenvalues} gives
\begin{equation}
  \lambda_+\simeq-\left(
  \frac{P_l}{\tau_{v,l}}+\frac{P_h}{\tau_{v,h}}
  \right),
  \quad
  \lambda_-=-K+O(1).
  \label{eq:lambda_Kinf}
\end{equation}
We introduce the effective velocity-relaxation time
\begin{equation}
  \frac{1}{\tau_{\rm eff}}
  \equiv
  \frac{P_l}{\tau_{v,l}}+\frac{P_h}{\tau_{v,h}}
  =\frac{\langle\gamma\rangle_{\rm eq}}{m}.
  \label{eq:tau_eff}
\end{equation}
Thus, the fast eigenmode $\lambda_-$ disappears on the timescale
$K^{-1}$, whereas the remaining mode relaxes with the finite timescale
$\tau_{\rm eff}$.

Taking $K\to\infty$ in the general finite-time result
Eq.~\eqref{eq:Dunder}, we obtain
\begin{equation}
  \lim_{K\to\infty}D_{\rm eff}^{\rm under}(t,K)
  =
  \frac{k_{\rm B}T\tau_{\rm eff}}{m}
  \left[
    1-\frac{\tau_{\rm eff}}{t}
    \left(1-e^{-t/\tau_{\rm eff}}\right)
  \right].
  \label{eq:Dunder_Kinf}
\end{equation}
This expression is identical to the finite-time effective diffusion
coefficient of a single-state underdamped Langevin dynamics with
constant friction $\langle\gamma\rangle_{\rm eq}$. We therefore define
\begin{equation}
  D_{\gamma}^{\rm single}(t)
  \equiv
  \frac{k_{\rm B}T}{\gamma}
  \left[
    1-\frac{\tau_v}{t}
    \left(1-e^{-t/\tau_v}\right)
  \right],
  \label{eq:D_single}
\end{equation}
and $\tau_v=m/\gamma$.
Since $\tau_{\rm eff}=m/\langle\gamma\rangle_{\rm eq}$,
Eq.~\eqref{eq:Dunder_Kinf} can be written compactly as
\begin{equation}
  \lim_{K\to\infty}D_{\rm eff}^{\rm under}(t,K)
  =
  D_{\langle\gamma\rangle_{\rm eq}}^{\rm single}(t).
  \label{eq:Dunder_Kinf_single}
\end{equation}
Thus, in the fast-switching limit, the finite-time diffusivity of the
underdamped model is the same as that of a single-state system with
friction $\langle\gamma\rangle_{\rm eq}$.

In contrast, the overdamped model gives
\begin{equation}
  \lim_{K\to\infty}D_{\rm eff}^{\rm over}(t,K)
  =
  k_{\rm B}T\langle\gamma^{-1}\rangle_{\rm eq}.
  \label{eq:Dover_Kinf}
\end{equation}
The difference between the two descriptions is therefore
\begin{align}
  \lim_{K\to\infty}\Delta D(t,K)
  ={}&
  k_{\rm B}T\left(
    \langle\gamma^{-1}\rangle_{\rm eq}
    -\frac{1}{\langle\gamma\rangle_{\rm eq}}
  \right)
  \notag\\
  &+
  \frac{k_{\rm B}T}{\langle\gamma\rangle_{\rm eq}}
  \frac{\tau_{\rm eff}}{t}
  \left(1-e^{-t/\tau_{\rm eff}}\right).
  \label{eq:DeltaD_Kinf}
\end{align}
The two terms in Eq.~\eqref{eq:DeltaD_Kinf} have distinct origins.
The second term is the usual finite-time inertial correction and
decays as $t^{-1}$ for $t\gg\tau_{\rm eff}$, whereas the first term
arises from the friction fluctuations and is independent of the
observation time. The latter is precisely the persistent long-time
discrepancy obtained in Sec.~\ref{sec:Deff} in the fast-switching limit.

\subsection{Slow-switching regime and two-step relaxation}
\label{sec:slow}

In the limit $K\to0$ at fixed $P_l$ and $P_h$, both switching
rates vanish. Accordingly,
\begin{equation}
  \lambda_+\to-\frac{1}{\tau_{v,l}},
  \quad
  \lambda_-\to-\frac{1}{\tau_{v,h}},
  \label{eq:lambda_K0}
\end{equation}
and
\begin{equation}
  A_+\to\frac{k_{\rm B}T}{m}p_l(0),
  \quad
  A_-\to\frac{k_{\rm B}T}{m}p_h(0).
\end{equation}
Taking the limit $K\to0$ in the general finite-time result then gives
\begin{equation}
  \lim_{K\to0}D_{\rm eff}^{\rm under}(t,K)
  =
  p_l(0)D_{\gamma_l}^{\rm single}(t)
  +
  p_h(0)D_{\gamma_h}^{\rm single}(t).
  \label{eq:D_K0_final}
\end{equation}
Equation~\eqref{eq:D_K0_final} has a simple physical interpretation.
In the limit $K\to0$, the environmental state remains frozen over any
finite observation time, so that each particle evolves as a single-state
underdamped Langevin system with its initial friction coefficient.
The ensemble-averaged diffusion coefficient is therefore a weighted
average of the two single-state results, with weights given by the
initial occupation probabilities $p_l(0)$ and $p_h(0)$.
Consequently, the finite-time diffusion coefficient retains explicit
memory of the initial environmental preparation in this limit.

\begin{figure}[t]
  \centering
  \includegraphics[width=0.95\linewidth]{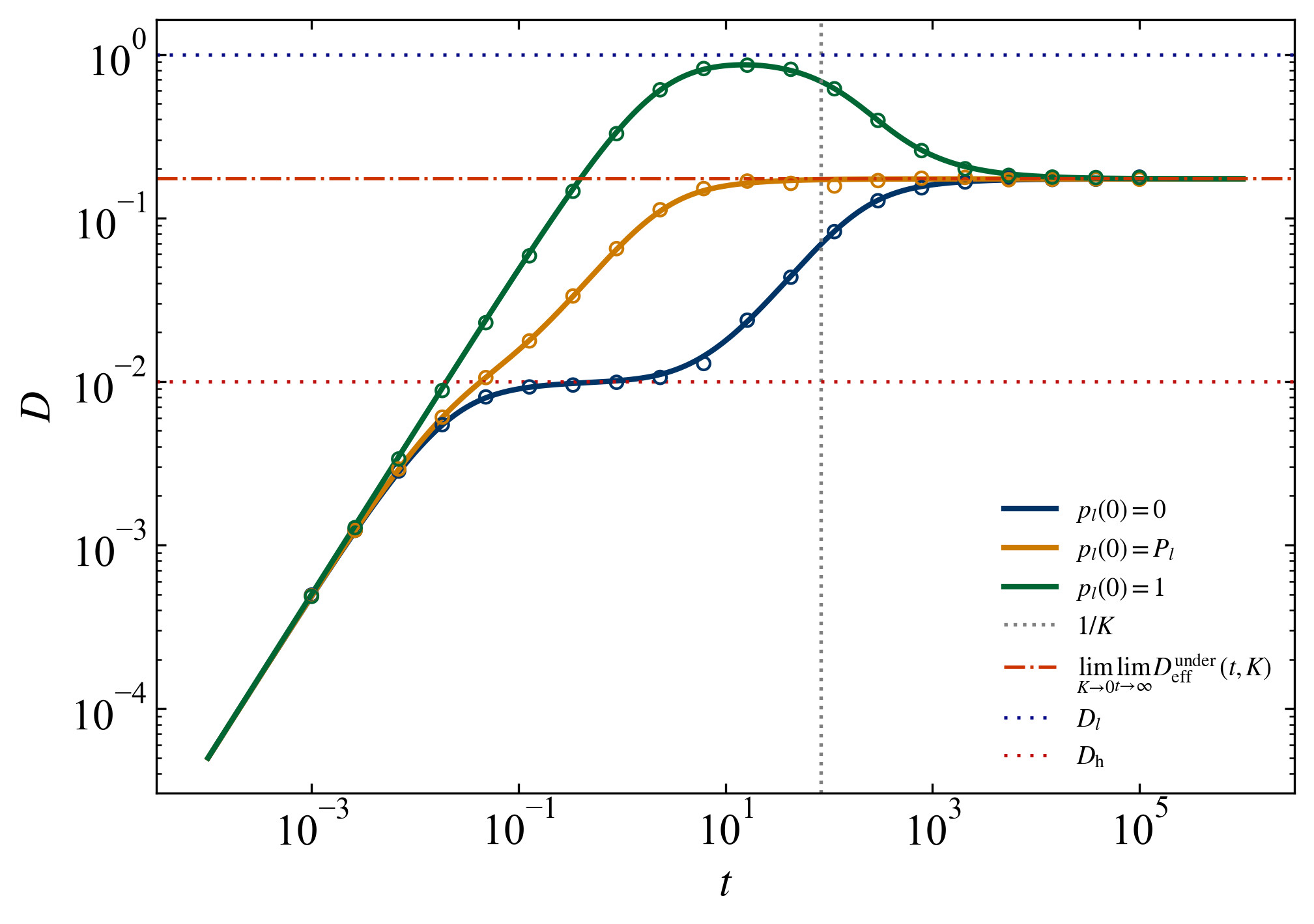}
  \caption{Finite-time underdamped effective diffusion coefficient
$D_{\rm eff}^{\rm under}(t,K)$ in the slow-switching regime for three
initial occupation probabilities, $p_l(0)=0$, $P_l$, and $1$.
The parameters are $\gamma_l=1$, $\gamma_h=100$, $m=k_{\rm B}T=1$,
$P_l=1/6$, $P_h=5/6$, $\tau_l=100$, $\tau_h=500$, and
$K=1.2\times10^{-2}$.
Solid lines show the theoretical results and symbols show simulations.
At times shorter than the environmental relaxation time $K^{-1}$,
the curves retain a strong dependence on the initial occupation and
approach the corresponding frozen-environment behavior predicted by
Eq.~\eqref{eq:D_K0_final}.
For $t\gtrsim K^{-1}$, environmental switching progressively erases
the initial-state dependence, and all curves converge toward the common
long-time value $D_{\rm eff}^{\rm under}(K)$.
The vertical dotted line indicates $t=K^{-1}$.
}
  \label{fig:Dunder_slow}
\end{figure}

For small but finite $K$, this initial-state memory persists on
timescales shorter than the environmental relaxation time $K^{-1}$.
When the velocity and environmental relaxation times are well separated,
$\tau_{v,l}\ll K^{-1}$, an intermediate time window
\begin{equation}
  \tau_{v,l}\ll t\ll K^{-1}
\end{equation}
emerges. In this regime, velocity relaxation is essentially complete,
whereas environmental switching remains rare, yielding the
initial-condition-dependent plateau
\begin{equation}
  D_{\rm eff}^{\rm under}(t,K)
  \simeq p_l(0)D_l+p_h(0)D_h.
  \label{eq:intermediate_plateau}
\end{equation}
At longer times, $t\gtrsim K^{-1}$, environmental switching erases
the memory of the initial occupation and drives the diffusion coefficient
toward its asymptotic value $D_{\rm eff}^{\rm under}(K)$.
Figure~\ref{fig:Dunder_slow} illustrates this two-step relaxation
for different initial environmental preparations.
At short times, the finite-time diffusion coefficient strongly depends
on $p_l(0)$, reflecting the initial environmental preparation.
After velocity relaxation, the curves approach the
initial-condition-dependent intermediate values predicted by
Eq.~\eqref{eq:intermediate_plateau}, which persist as long as
environmental switching remains rare.
Around the environmental relaxation time $t\sim K^{-1}$, the second
crossover sets in: switching progressively erases the memory of the
initial preparation, and the curves merge toward the same long-time
diffusion coefficient $D_{\rm eff}^{\rm under}(K)$.
The agreement between the theoretical curves and the simulation results
confirms the separation of the velocity- and environmental-relaxation
timescales underlying the two-step behavior.

The overdamped model provides a useful comparison. For
$t\ll K^{-1}$, environmental switching is negligible, and
Eq.~\eqref{eq:Dover_general} gives
\begin{equation}
  D_{\rm eff}^{\rm over}(t,K)
  \simeq p_l(0)D_l+p_h(0)D_h.
  \label{eq:Dover_slow_intermediate}
\end{equation}
Thus, after velocity relaxation but before appreciable environmental
switching, the two descriptions share the same intermediate plateau,
\begin{equation}
  D_{\rm eff}^{\rm under}(t,K)
  \simeq
  D_{\rm eff}^{\rm over}(t,K)
  \simeq
  p_l(0)D_l+p_h(0)D_h,
  \label{eq:common_intermediate_plateau}
\end{equation}
for $\tau_{v,l}\ll t\ll K^{-1}$.
At shorter times, the two descriptions differ because of the finite
velocity-relaxation time in the underdamped dynamics, whereas at longer
times they approach their respective long-time diffusion coefficients.
For sufficiently slow switching, the difference between the latter
vanishes as $K\to0$, consistently with Sec.~\ref{sec:Deff}.

The slow-switching regime therefore exhibits a two-step relaxation:
velocity relaxation first establishes an initial-condition-dependent
intermediate plateau, while environmental relaxation subsequently
erases this memory and drives the system toward its long-time
diffusive state.

\section{General fluctuating friction}
\label{sec:general}

We now return to the general fluctuating-friction dynamics introduced
in Sec.~\ref{sec: model} and show, under a standard fast-mixing
assumption, that the mechanism identified exactly for the two-state
model extends beyond dichotomous switching.

We consider a stationary positive stochastic process $\gamma(t)$ with
finite mean $\langle\gamma\rangle_{\rm eq}$ and
$\langle\gamma^{-1}\rangle_{\rm eq}$. We assume that $\gamma(t)$ is
statistically independent of the thermal noise and that its temporal
correlations decay on a characteristic environmental timescale
$\tau_{\rm env}$. The underdamped and overdamped dynamics are those
defined in Eqs.~\eqref{eq:ULE} and~\eqref{eq:OD_model}, respectively.

For the overdamped dynamics, the mean-squared displacement conditioned
on a realization of $\gamma(t)$ is
\begin{equation}
  \left\langle [x(t)-x(0)]^2\mid\gamma\right\rangle
  =
  2k_{\rm B}T
  \int_0^t\frac{ds}{\gamma(s)}.
  \label{eq:general_OD_conditional_MSD}
\end{equation}
Averaging over a stationary friction process therefore gives
\begin{equation}
  D_{\rm eff}^{\rm over}
  =
  k_{\rm B}T
  \left\langle\gamma^{-1}\right\rangle_{\rm eq}.
  \label{eq:general_Dover}
\end{equation}
Thus, irrespective of the temporal correlations of the friction,
the overdamped diffusion coefficient is governed by the mean inverse
friction.

The underdamped dynamics behaves differently when the environmental
fluctuations become rapid compared with velocity relaxation. To make
this limit explicit, consider a family of stationary friction processes
$\gamma_\epsilon(t)=\gamma(t/\epsilon)$, whose correlation time decreases
as $\epsilon\to0$ while their stationary distribution is kept fixed.
The velocity equation then reads
\begin{equation}
  m\dot v(t)
  =
  -\gamma_\epsilon(t)v(t)
  +\sqrt{2k_{\rm B}T\gamma_\epsilon(t)}\,\xi(t).
  \label{eq:general_ULE_fast}
\end{equation}
On timescales over which $v(t)$ changes appreciably, the rapidly
fluctuating friction samples its stationary distribution many times.
Consequently, the integrated friction self-averages,
\begin{equation}
  \int_0^t\gamma_\epsilon(s)\,ds
  \xrightarrow[\epsilon\to0]{}
  t\langle\gamma\rangle_{\rm eq},
  \label{eq:general_gamma_average}
\end{equation}
and the quadratic variation of the noise term satisfies
\begin{equation}
  \int_0^t 2k_{\rm B}T\gamma_\epsilon(s)\,ds
  \xrightarrow[\epsilon\to0]{}
  2k_{\rm B}T\langle\gamma\rangle_{\rm eq}t.
  \label{eq:general_noise_average}
\end{equation}
The limiting velocity dynamics is therefore
\begin{equation}
  m\dot v(t)
  =
  -\langle\gamma\rangle_{\rm eq}v(t)
  +\sqrt{2k_{\rm B}T\langle\gamma\rangle_{\rm eq}}\,\xi(t),
  \label{eq:general_fast_ULE}
\end{equation}
which is the single-friction underdamped Langevin dynamics with the
arithmetic mean friction. Its long-time diffusion coefficient is
\begin{equation}
  \lim_{\tau_{\rm env}\to0}
  D_{\rm eff}^{\rm under}
  =
  \frac{k_{\rm B}T}{\langle\gamma\rangle_{\rm eq}}.
  \label{eq:general_Dunder_fast}
\end{equation}

Combining Eqs.~\eqref{eq:general_Dover} and
\eqref{eq:general_Dunder_fast}, we obtain
\begin{equation}
  \lim_{\tau_{\rm env}\to0}
  \left(
  D_{\rm eff}^{\rm over}
  -
  D_{\rm eff}^{\rm under}
  \right)
  =
  k_{\rm B}T
  \left(
  \langle\gamma^{-1}\rangle_{\rm eq}
  -
  \frac{1}{\langle\gamma\rangle_{\rm eq}}
  \right).
  \label{eq:general_fast_gap}
\end{equation}
Since
\begin{equation}
  \langle\gamma^{-1}\rangle_{\rm eq}
  \geq
  \frac{1}{\langle\gamma\rangle_{\rm eq}},
  \label{eq:Jensen_general}
\end{equation}
with equality only for a nonfluctuating friction, the overdamped model
generically overestimates the long-time diffusion coefficient in the
fast-fluctuation limit.

The corresponding relative error is
\begin{equation}
  \varepsilon_{\rm OD}^{\rm fast}
  =
  1-
  \frac{1}{
  \langle\gamma\rangle_{\rm eq}
  \langle\gamma^{-1}\rangle_{\rm eq}},
  \label{eq:general_eps_fast}
\end{equation}
which has exactly the same form as the saturation value obtained for
the two-state model in Eq.~\eqref{eq:eps_fast}. Hence, the persistent
breakdown of the overdamped approximation is not a consequence of
dichotomous switching. It originates from the different averaging
operations performed by the two descriptions: rapid environmental
fluctuations lead the underdamped velocity to respond to the arithmetic
mean friction $\langle\gamma\rangle_{\rm eq}$, whereas the overdamped
dynamics averages the instantaneous mobility and is therefore governed
by $\langle\gamma^{-1}\rangle_{\rm eq}$.

To test the general fast-fluctuation prediction for a continuous
environmental process, we consider a log-Ornstein--Uhlenbeck friction.
Taking $\gamma(t)$ itself to be an Ornstein--Uhlenbeck process is not
admissible here, since it can attain negative values and
$\langle\gamma^{-1}\rangle_{\rm eq}$ need not exist. 
We therefore introduce an Ornstein--Uhlenbeck process for the
logarithmic friction,
\begin{equation}
  \dot{Y}(t)
  =
  -\frac{Y(t)-\mu}{\tau_{\rm env}}
  +\sqrt{\frac{2s^{2}}{\tau_{\rm env}}}\,\xi_Y(t),
\end{equation}
where $\mu$ is the stationary mean of $Y(t)$,
$\tau_{\rm env}$ is its relaxation time, and $\xi_Y(t)$ is Gaussian
white noise independent of the thermal noise $\xi(t)$, satisfying
$\langle\xi_Y(t)\rangle=0$ and
$\langle\xi_Y(t)\xi_Y(t')\rangle=\delta(t-t')$.
Defining $\gamma(t)=e^{Y(t)}$ guarantees $\gamma(t)>0$ for all $t$
and ensures that both $\langle\gamma\rangle_{\rm eq}$ and
$\langle\gamma^{-1}\rangle_{\rm eq}$ are finite.
The stationary distribution of $Y$ is Gaussian,
$Y\sim\mathcal{N}(\mu,s^{2})$, with
$s^{2}=\mathrm{Var}[\ln\gamma]$.

A useful feature of this parametrization is that the stationary distribution
of $\gamma$ is independent of $\tau_{\rm env}$, so that varying
$\tau_{\rm env}$ changes the environmental switching speed alone while
leaving the friction statistics untouched. This is the direct analogue of
holding $P_{l}$, $P_{h}$, $\gamma_{l}$, $\gamma_{h}$ fixed and varying $K$ in
the two-state model of Sec.~\ref{sec:two_state}.

The stationary moments follow in closed form,
\begin{equation}
  \langle\gamma\rangle_{\rm eq} = e^{\mu+s^{2}/2},
  \quad
  \langle\gamma^{-1}\rangle_{\rm eq} = e^{-\mu+s^{2}/2},
\end{equation}
so that the corresponding diffusion coefficients are
\begin{equation}
  \begin{aligned}
    \lim_{\tau_{\rm env}\to0} D_{\rm eff}^{\rm under}
      &= \frac{k_{\rm B}T}{\langle\gamma\rangle_{\rm eq}}
      = k_{\rm B}T\,e^{-\mu-s^{2}/2},\\
    D_{\rm eff}^{\rm over}
      &= k_{\rm B}T\,\langle\gamma^{-1}\rangle_{\rm eq}
      = k_{\rm B}T\,e^{-\mu+s^{2}/2}.
  \end{aligned}
\end{equation}
Their product satisfies
$\langle\gamma\rangle_{\rm eq}\langle\gamma^{-1}\rangle_{\rm eq}=e^{s^{2}}$,
which depends only on the variance of $\ln\gamma$ and not on its mean, so
Eq.~\eqref{eq:general_eps_fast} reduces to
\begin{equation}
  \varepsilon_{\rm OD}^{\rm fast} = 1 - e^{-s^{2}} .
  \label{eq: error_fast}
\end{equation}
The independence of $\mu$ allows us to set $\mu=0$ without loss of
generality in what follows.

\begin{figure*}[t]
  \centering
  \includegraphics[width=0.48\textwidth]{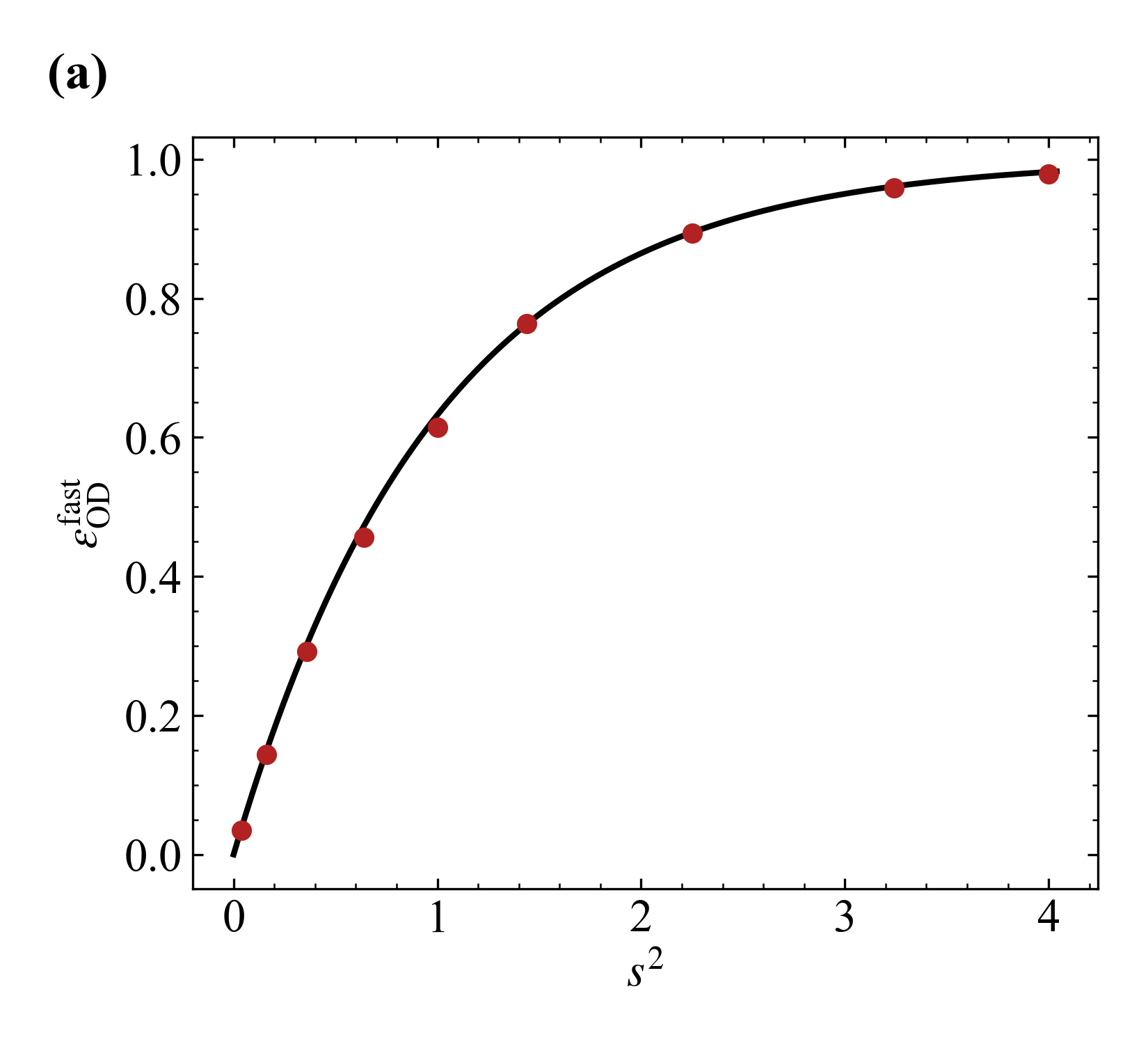}\hfill
  \includegraphics[width=0.48\textwidth]{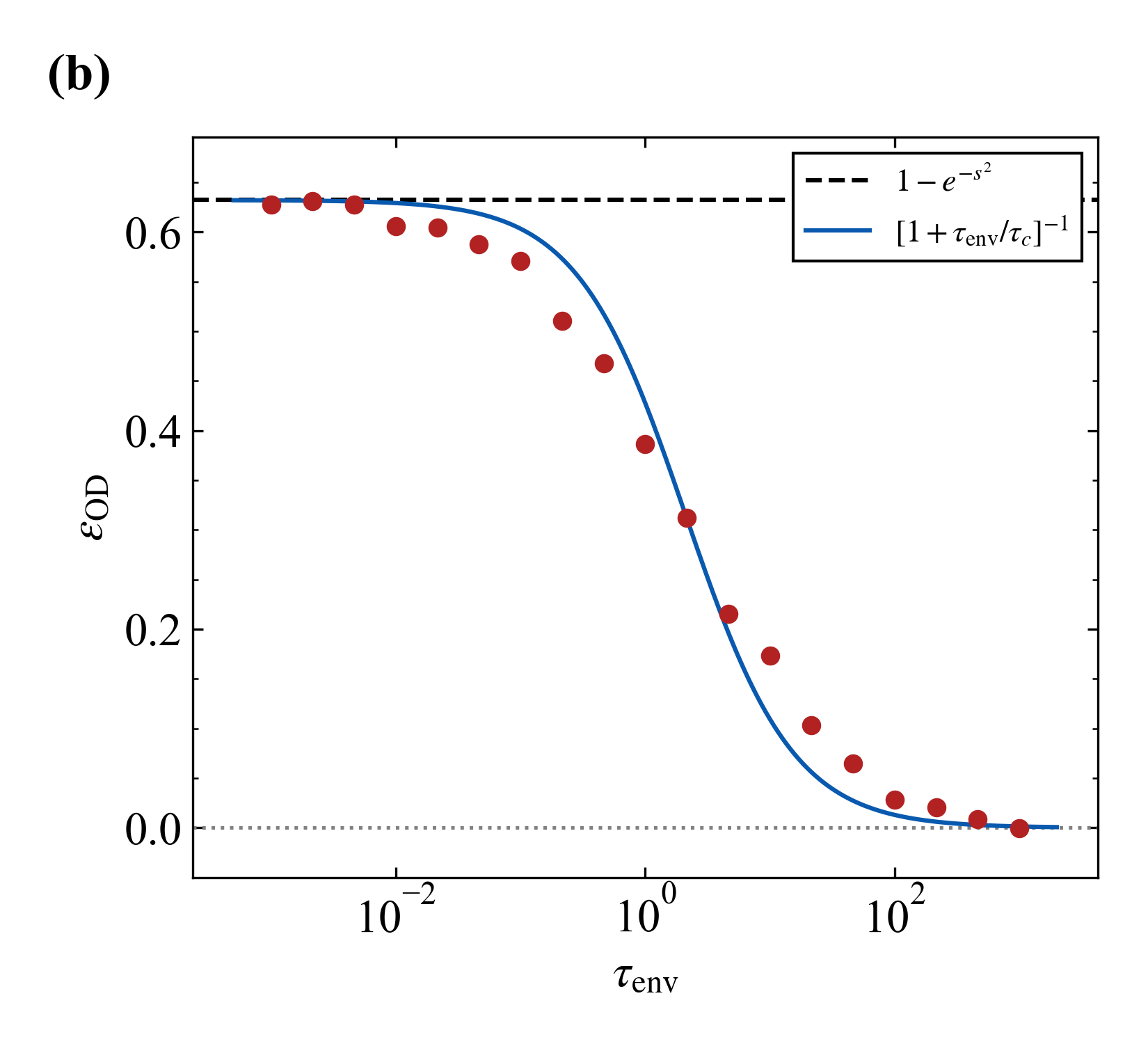}
  \caption{
(a) Fast-fluctuation relative error
$\varepsilon_{\rm OD}^{\rm fast}$ for the log-Ornstein--Uhlenbeck
friction as a function of the variance
$s^2=\mathrm{Var}[\ln\gamma]$ ($\mu=0$ and
$\tau_{\rm env}=10^{-3}$).
The solid line shows the general fast-fluctuation prediction
$\varepsilon_{\rm OD}^{\rm fast}=1-e^{-s^2}$ from Eq.~\eqref{eq: error_fast}, and
symbols show numerical simulations.
The agreement demonstrates that the persistent overdamped discrepancy
is not specific to dichotomous switching.
(b) Crossover of the relative error $\varepsilon_{\rm OD}$ with the
environmental timescale $\tau_{\rm env}$ for the
log-Ornstein--Uhlenbeck friction at fixed $s=1$ and $\mu=0$.
As $\tau_{\rm env}\to\infty$, the velocity has sufficient time to
relax within each frictional environment, so that both descriptions
reduce to $k_{\rm B}T\langle\gamma^{-1}\rangle_{\rm eq}$ and
$\varepsilon_{\rm OD}$ vanishes (dotted line).
The solid line shows the two-state crossover form
$\varepsilon_{\rm OD}^{\rm fast}
[1+\tau_{\rm env}/\tau_c]^{-1}$ of
Eq.~\eqref{eq:eps_crossover}, evaluated using the independently
determined crossover time $\tau_c=\tau_c^{\rm th}=2.094$.
No fitting parameter is used.
}
  \label{fig:general_fast_error}
\end{figure*}

Figure~\ref{fig:general_fast_error}(a) illustrates this
fast-fluctuation relative error as a function of the variance parameter
$s^2$ of the fluctuating-friction distribution.
The agreement between the general fast-mixing prediction and the
log-Ornstein--Uhlenbeck simulations shows that the breakdown is not
a peculiarity of dichotomous switching, but a consequence of temporal
friction fluctuations themselves.

Figure~\ref{fig:general_fast_error}(b) shows the
crossover of the relative error as the environmental correlation time
$\tau_{\rm env}$ is varied at fixed $s=1$.
For $\tau_{\rm env}\to0$, the friction is rapidly mixing and the
numerical values saturate at the fast-fluctuation prediction
$1-e^{-s^2}=0.632$ of Eq.~(92) (dashed line).
In the opposite limit $\tau_{\rm env}\to\infty$, the velocity has
sufficient time to relax within each frictional environment before the
environment changes, so that the overdamped result is recovered and
$\varepsilon_{\rm OD}$ vanishes.
Between these limits, we characterize the crossover by the
environmental timescale $\tau_c$ at which $\varepsilon_{\rm OD}$
reaches one half of its fast-fluctuation value.
As shown in Appendix~\ref{app:logOU_crossover}, the underdamped
diffusion coefficient obeys the scaling form
\begin{equation}
D_{\rm eff}^{\rm under}
=
k_{\rm B}T e^{-\mu}
{\cal D}\left(
\frac{\tau_{\rm env}e^\mu}{m},s^2
\right),
\end{equation}
which gives
\begin{equation}
\tau_c=m e^{-\mu}f(s^2),
\end{equation}
where ${\cal D}$ and $f$ are defined in
Appendix~\ref{app:logOU_crossover}.
For $m=1$, $\mu=0$, and $s^2=1$, we obtain
$\tau_c=2.094$.
The blue solid line shows the two-state crossover form
\begin{equation}
\varepsilon_{\rm OD}
=
\varepsilon_{\rm OD}^{\rm fast}
\left(1+\frac{\tau_{\rm env}}{\tau_c}\right)^{-1},
\end{equation}
derived for dichotomous switching, with the independently determined
value $\tau_c=2.094$ and no fitting parameters.
Remarkably, this expression captures both the location and the overall
shape of the crossover for the log-Ornstein--Uhlenbeck friction,
although the numerical data decay somewhat more slowly on the
large-$\tau_{\rm env}$ side.
Thus, while the detailed crossover shape depends on the temporal
statistics of the environment, the two-state crossover form provides
a useful description of the crossover once its characteristic
timescale is determined from the underlying environmental dynamics.

\section{Discussion}

The two-state friction model provides a minimal dynamical representation
of temporal heterogeneity encountered by probe particles in glass-forming
liquids, supercooled liquids, and other complex media. In such systems,
different local environments may be associated not only with different
instantaneous diffusivities but also with different velocity-relaxation
times. Our results show that this additional dynamical timescale can be
essential for assessing the validity of an overdamped description.

This observation refines the usual criterion for an overdamped
description. In a static environment, it is sufficient for the observation
time to be long compared with the velocity-relaxation time. In a
fluctuating environment, however, velocity relaxation must also be fast
compared with environmental switching. For the two-state model, this
competition is quantified by
\begin{equation}
  \Theta
  =
  \frac{\tau_{v,l}}{\tau_l}
  +
  \frac{\tau_{v,h}}{\tau_h}.
\end{equation}
When $\Theta\ll1$, the velocity can relax within each environment before
the next transition, and the overdamped approximation becomes accurate
at long times. When $\Theta$ is of order unity or larger, inertial memory
can persist across environmental transitions. This condition is sufficient
but not necessary for practical accuracy: the relative error also depends
on the friction heterogeneity through $\varepsilon_{\rm OD}^{\rm fast}$
and can remain small for $\Theta\gtrsim1$ when the friction contrast is
weak.

The present model is closely related to fluctuating-diffusivity
descriptions, in which a stochastic environmental variable modulates the
local diffusion coefficient. The overdamped model considered here belongs
to this class, with $D(t)=k_{\rm B}T/\gamma(t)$. Our results show,
however, that the statistics of the fluctuating diffusivity alone do not
generally determine the transport of a particle with finite inertia.
The velocity retains information about the history of the environmental
state that is absent from a purely overdamped description. In particular,
rapid switching exposes the noncommutativity directly: the underdamped
dynamics responds to the arithmetic mean friction
$\langle\gamma\rangle_{\rm eq}$, whereas the overdamped dynamics is
governed by the mean inverse friction
$\langle\gamma^{-1}\rangle_{\rm eq}$.

Finite observation time provides another manifestation of the interplay
between the two relaxation processes. In the slow-switching regime,
velocity and environmental relaxation become well separated, giving rise
to the two-step behavior discussed in Sec.~\ref{sec:slow}. The system
first approaches an initial-condition-dependent intermediate plateau,
while environmental switching subsequently erases this memory on the
timescale $K^{-1}$. Consequently, diffusion coefficients inferred from
finite trajectories can depend on both the observation window and the
initial environmental preparation.

More generally, eliminating the fluctuating environmental variable does
not necessarily yield a Langevin equation with a single renormalized
friction coefficient. At finite switching rate, the two relaxation
eigenmodes $\lambda_\pm$ reflect memory generated by the unresolved
environmental state. A reduced description involving only position or
velocity can therefore involve memory, correlated effective noise, or
additional hidden variables.

The consequences of this noncommutativity may extend beyond diffusion
to stochastic thermodynamics. Inertia is known to influence the
performance of stochastic heat engines
\cite{Tu2014,dechant2017underdamped,Nakamura2020,Fora02024,p2026inertia},
and exact finite-time fluctuation--response relations and associated
thermodynamic bounds have recently been established for underdamped
Langevin dynamics \cite{van2026exact}. If the system--environment
coupling itself fluctuates on timescales comparable to velocity
relaxation, eliminating inertia before averaging over these fluctuations
may therefore affect not only transport coefficients but also response,
work, and heat currents. This raises the broader question of whether
environmental averaging and inertial elimination lead to distinct
thermodynamic descriptions.

\section{Conclusions}

We have investigated the validity of the overdamped approximation
for Brownian dynamics in temporally fluctuating frictional
environments. For a two-state Markov friction, we derived the
finite-time effective diffusion coefficient exactly and showed that
the validity of the overdamped approximation is controlled by the
competition between velocity relaxation and environmental switching.

The long-time discrepancy between the underdamped and overdamped
descriptions is characterized by the dimensionless parameter
$\Theta=\tau_{v,l}/\tau_l+\tau_{v,h}/\tau_h$ and by the friction
heterogeneity. For slow switching, the velocity relaxes within each
environmental state and the overdamped result is recovered. For fast
switching, by contrast, the underdamped dynamics is governed by the
arithmetic mean friction $\langle\gamma\rangle_{\rm eq}$, whereas the
overdamped dynamics is governed by the mean inverse friction
$\langle\gamma^{-1}\rangle_{\rm eq}$. The resulting discrepancy persists
even at infinite observation time, with the fast-switching relative error
\begin{equation}
  \varepsilon_{\rm OD}^{\rm fast}
  =
  1-
  \frac{1}{
  \langle\gamma\rangle_{\rm eq}
  \langle\gamma^{-1}\rangle_{\rm eq}}.
\end{equation}

At finite observation times, slow environmental switching produces a
two-step relaxation: velocity relaxation first establishes an
initial-condition-dependent intermediate plateau, while environmental
relaxation subsequently erases this memory and drives the system toward
its asymptotic diffusive state. Thus, a separation between the observation
time and the conventional inertial timescale alone is not sufficient to
justify an overdamped description in a temporally heterogeneous
environment; the environmental relaxation timescale must also be taken
into account.

More generally, we showed that the fast-fluctuation discrepancy is not
specific to dichotomous switching. For a broad class of rapidly mixing
friction processes, the underdamped and overdamped descriptions converge
to diffusion coefficients governed by $\langle\gamma\rangle_{\rm eq}$ and
$\langle\gamma^{-1}\rangle_{\rm eq}$, respectively. Numerical results for
a log-Ornstein--Uhlenbeck friction process confirm this general prediction
for a continuous fluctuating environment. The breakdown of the overdamped
approximation therefore reflects a noncommutativity between environmental
averaging and inertial elimination, rather than a peculiarity of the
two-state model.

Our results show that rapid environmental fluctuations do not necessarily
wash out inertial effects. On the contrary, they can leave a persistent
signature in long-time transport even when the observation time is much
longer than the velocity-relaxation time. This provides a general
criterion for assessing overdamped descriptions in dynamically
heterogeneous environments.

\appendix

%%% APPENDIX CANDIDATE (begin): short-time path decomposition for C_w

\section{Short-time derivation of the correlation equations}
\label{app: Expansion}

We derive the evolution equations for $C_l$ and $C_h$ by decomposing
the possible paths over a short lag-time increment $ds$. For brevity,
let $u=t_1+s$. The It\^o increment of the velocity is
\begin{equation}
  v(u+ds)
  =
  v(u)\left(1-\frac{\gamma_{\sigma(u)}}{m}ds\right)
  +
  \sqrt{\frac{2k_{\rm B}T\gamma_{\sigma(u)}}{m^2}}\,dW(u).
  \label{eq:v_increment}
\end{equation}
Since $dW(u)$ is independent of the past for $u\ge t_1$,
$\langle v(t_1)dW(u)\rangle=0$. In addition, the environmental
switching occurs with constant rates $k_l$ and $k_h$, independently
of the velocity. These properties lead to a closed evolution equation
for the two state-resolved correlations.

We first consider paths that remain in the low-friction state during
$[u,u+ds]$. Their contribution is
\begin{align}
  &\left\langle
  v(t_1)v(u+ds)
  \mathbf{1}_{\{\sigma(u)=l,\,\sigma(u+ds)=l\}}
  \right\rangle
  \notag\\
  &\qquad=
  (1-k_l ds)
  \left(1-\frac{\gamma_l}{m}ds\right)
  C_l(t_1,u)+o(ds).
  \label{eq:Cl_ll}
\end{align}
Here $1-k_l ds$ is the probability that no $l\to h$ transition
occurs during the increment, while the second factor describes
velocity relaxation in state $l$.

Paths that switch from the high-friction state to the low-friction
state during the same increment contribute
\begin{align}
  &\left\langle
  v(t_1)v(u+ds)
  \mathbf{1}_{\{\sigma(u)=h,\,\sigma(u+ds)=l\}}
  \right\rangle
  \notag\\
  &\qquad=
  k_h ds\,C_h(t_1,u)+o(ds).
  \label{eq:Cl_hl}
\end{align}
The velocity remains continuous at the switching event. Since the
switching probability is already of order $ds$, corrections due to
velocity relaxation during such a path are of order $ds^2$.

Combining the two contributions gives
\begin{align}
  C_l(t_1,u+ds)
  ={}&
  (1-k_l ds)
  \left(1-\frac{\gamma_l}{m}ds\right)
  C_l(t_1,u)
  \notag\\
  &+k_h ds\,C_h(t_1,u)+o(ds).
\end{align}
Using $u=t_1+s$ and taking $ds\to0$, we obtain
\begin{equation}
  \frac{\partial C_l}{\partial s}
  =
  -\left(\frac{\gamma_l}{m}+k_l\right)C_l
  +k_h C_h.
  \label{eq:app_dCl}
\end{equation}

An analogous decomposition for paths ending in the high-friction
state yields
\begin{equation}
  \frac{\partial C_h}{\partial s}
  =
  k_l C_l
  -\left(\frac{\gamma_h}{m}+k_h\right)C_h.
  \label{eq:app_dCh}
\end{equation}
Thus, velocity relaxation contributes the diagonal decay rates
$\gamma_w/m$, while environmental transitions transfer correlation
between the two state-resolved components with rates $k_l$ and $k_h$.

\section{Derivation of the state-resolved second moments}
\label{app: Phi}

We derive the evolution equations for the state-resolved second moments
$\Phi_l(t)$ and $\Phi_h(t)$. Over a short time increment $dt$,
$\Phi_l(t+dt)$ receives contributions from paths that remain in state $l$
and from paths that switch from $h$ to $l$:
\begin{align}
  \Phi_l(t+dt)
  ={}&
  \left\langle
  v^2(t+dt)
  \mathbf{1}_{\{\sigma(t)=l,\,\sigma(t+dt)=l\}}
  \right\rangle
  \notag\\
  &+
  \left\langle
  v^2(t+dt)
  \mathbf{1}_{\{\sigma(t)=h,\,\sigma(t+dt)=l\}}
  \right\rangle .
  \label{eq:Phi_l_decomposition}
\end{align}
For paths that remain in state $l$, It\^o's formula for $v^2$ gives
\begin{align}
  &\left\langle
  v^2(t+dt)
  \mathbf{1}_{\{\sigma(t)=l,\,\sigma(t+dt)=l\}}
  \right\rangle
  \notag\\
  &=
  \Phi_l(t)
  -\left(\frac{2\gamma_l}{m}+k_l\right)\Phi_l(t)\,dt
  +\frac{2k_{\rm B}T\gamma_l}{m^2}p_l(t)\,dt
  +o(dt).
  \label{eq:Phi_ll}
\end{align}
For paths that switch from $h$ to $l$, the switching probability is
$k_hdt$ and the velocity remains continuous at the switching event.
Hence,
\begin{equation}
  \left\langle
  v^2(t+dt)
  \mathbf{1}_{\{\sigma(t)=h,\,\sigma(t+dt)=l\}}
  \right\rangle
  =
  k_h\Phi_h(t)\,dt+o(dt).
  \label{eq:Phi_hl}
\end{equation}
Combining these contributions and taking $dt\to0$ yields
\begin{equation}
  \frac{d\Phi_l}{dt}
  =
  -\left(\frac{2\gamma_l}{m}+k_l\right)\Phi_l
  +k_h\Phi_h
  +\frac{2k_{\rm B}T\gamma_l}{m^2}p_l.
  \label{eq:dPhif}
\end{equation}
Similarly,
\begin{equation}
  \frac{d\Phi_h}{dt}
  =
  k_l\Phi_l
  -\left(\frac{2\gamma_h}{m}+k_h\right)\Phi_h
  +\frac{2k_{\rm B}T\gamma_h}{m^2}p_h.
  \label{eq:dPhis}
\end{equation}

To show that these equations preserve equipartition within each
environmental state, define
\begin{equation}
  \Psi_w(t)
  \equiv
  \Phi_w(t)-\frac{k_{\rm B}T}{m}p_w(t).
  \label{eq:Psi_def}
\end{equation}
Using Eqs.~\eqref{eq:dPhif}--\eqref{eq:dPhis} and the
occupation-probability equations
\begin{equation}
  \dot p_l=-k_l p_l+k_h p_h,
  \qquad
  \dot p_h=k_l p_l-k_h p_h,
  \label{eq:p_master}
\end{equation}
we obtain
\begin{equation}
  \frac{d}{dt}
  \begin{pmatrix}
    \Psi_l\\
    \Psi_h
  \end{pmatrix}
  =
  \begin{pmatrix}
    -2\gamma_l/m-k_l & k_h\\
    k_l & -2\gamma_h/m-k_h
  \end{pmatrix}
  \begin{pmatrix}
    \Psi_l\\
    \Psi_h
  \end{pmatrix}.
  \label{eq:Psi_matrix}
\end{equation}

For a Maxwellian initial velocity statistically independent of
$\sigma(0)$, $\Psi_l(0)=\Psi_h(0)=0$. Since
Eq.~\eqref{eq:Psi_matrix} is homogeneous, it follows immediately that
$\Psi_l(t)=\Psi_h(t)=0$ for all $t\ge0$. Hence,
\begin{equation}
  \Phi_w(t)=\frac{k_{\rm B}T}{m}p_w(t).
\end{equation}

\section{Crossover time for log-Ornstein--Uhlenbeck friction}
\label{app:logOU_crossover}

Here we determine the characteristic crossover time for the
log-Ornstein--Uhlenbeck friction introduced in Sec.~\ref{sec:general}. The logarithm
of the friction coefficient obeys
\begin{equation}
  \dot Y
  =
  -\frac{Y-\mu}{\tau_{\rm env}}
  +
  \sqrt{\frac{2s^2}{\tau_{\rm env}}}\,\xi_Y(t),
  \qquad
  \gamma(t)=e^{Y(t)},
\end{equation}
where $\xi_Y(t)$ is Gaussian white noise. In the stationary state,
$Y$ is normally distributed with mean $\mu$ and variance $s^2$.

For a stationary friction process independent of the thermal noise,
the velocity autocorrelation function can be written as
\begin{equation}
  C_v(t)
  =
  \frac{k_{\rm B}T}{m}
  \left\langle
  \exp\left[
  -\frac{1}{m}\int_0^t \gamma(u)\,du
  \right]
  \right\rangle_{\rm eq}.
  \label{eq:logOU_Cv}
\end{equation}
The long-time underdamped diffusion coefficient therefore follows
from the Green--Kubo relation as
\begin{equation}
  D_{\rm eff}^{\rm under}(\tau_{\rm env})
  =
  \frac{k_{\rm B}T}{m}
  \int_0^\infty
  \left\langle
  \exp\left[
  -\frac{1}{m}\int_0^t e^{Y(u)}\,du
  \right]
  \right\rangle_{\rm eq}
  dt.
  \label{eq:logOU_GK}
\end{equation}

The dependence on $m$ and $\mu$ can be separated by introducing
\begin{equation}
  Y(t)=\mu+s Z(t)
\end{equation}
and the rescaled time
\begin{equation}
  u=\frac{e^\mu}{m}t.
\end{equation}
In terms of $u$, the stationary process $Z$ obeys
\begin{equation}
  \frac{dZ}{du}
  =
  -\frac{Z}{r}
  +
  \sqrt{\frac{2}{r}}\,\xi_Z(u),
  \qquad
  r\equiv\frac{\tau_{\rm env}e^\mu}{m},
  \label{eq:logOU_r}
\end{equation}
where $Z$ has zero mean and unit variance. Equation~\eqref{eq:logOU_GK}
then takes the scaling form
\begin{equation}
  D_{\rm eff}^{\rm under}
  =
  k_{\rm B}T e^{-\mu}
  {\cal D}(r,s^2),
  \label{eq:logOU_scaling}
\end{equation}
where
\begin{equation}
  {\cal D}(r,s^2)
  =
  \int_0^\infty
  \left\langle
  \exp\left[
  -\int_0^u e^{sZ(u')}\,du'
  \right]
  \right\rangle_{\rm eq}
  du.
  \label{eq:logOU_Dcal}
\end{equation}
Thus, apart from the overall factor $k_{\rm B}T e^{-\mu}$,
the dependence on the environmental timescale enters only through
the dimensionless combination $r=\tau_{\rm env}e^\mu/m$.

For the log-Ornstein--Uhlenbeck process,
\begin{equation}
  \langle\gamma\rangle_{\rm eq}
  =e^{\mu+s^2/2},
  \qquad
  \langle\gamma^{-1}\rangle_{\rm eq}
  =e^{-\mu+s^2/2}.
\end{equation}
The fast- and slow-environment limits of the underdamped diffusion
coefficient are therefore
\begin{align}
  D_{\rm eff}^{\rm under,fast}
  &=
  \frac{k_{\rm B}T}{\langle\gamma\rangle_{\rm eq}}
  =
  k_{\rm B}T e^{-\mu-s^2/2},
  \\
  D_{\rm eff}^{\rm under,slow}
  &=
  k_{\rm B}T
  \langle\gamma^{-1}\rangle_{\rm eq}
  =
  k_{\rm B}T e^{-\mu+s^2/2}.
\end{align}
The latter coincides with the overdamped diffusion coefficient.

We define the crossover time $\tau_c$ as the value of
$\tau_{\rm env}$ at which the relative overdamped error reaches
one half of its fast-fluctuation value,
\begin{equation}
  \varepsilon_{\rm OD}(\tau_c)
  =
  \frac{1}{2}\varepsilon_{\rm OD}^{\rm fast}.
  \label{eq:logOU_midpoint}
\end{equation}
Using
\begin{equation}
  \varepsilon_{\rm OD}
  =
  1-
  \frac{D_{\rm eff}^{\rm under}}
       {D_{\rm eff}^{\rm over}},
\end{equation}
Eq.~\eqref{eq:logOU_midpoint} is equivalent to
\begin{equation}
  D_{\rm eff}^{\rm under}(\tau_c)
  =
  \frac{1}{2}
  \left(
  D_{\rm eff}^{\rm under,fast}
  +
  D_{\rm eff}^{\rm over}
  \right).
  \label{eq:logOU_midpoint_D}
\end{equation}
For the log-Ornstein--Uhlenbeck friction, this condition becomes
\begin{equation}
  D_{\rm eff}^{\rm under}(\tau_c)
  =
  k_{\rm B}T e^{-\mu}
  \cosh\left(\frac{s^2}{2}\right).
  \label{eq:logOU_midpoint_explicit}
\end{equation}

Combining Eqs.~\eqref{eq:logOU_scaling} and
\eqref{eq:logOU_midpoint_explicit}, we define $f(s^2)$ implicitly by
\begin{equation}
  {\cal D}\left(f(s^2),s^2\right)
  =
  \cosh\left(\frac{s^2}{2}\right).
  \label{eq:logOU_f}
\end{equation}
The crossover time then takes the scaling form
\begin{equation}
  \tau_c
  =
  m e^{-\mu}f(s^2).
  \label{eq:logOU_tauc}
\end{equation}
Thus, the dependence of the crossover time on the mass and the mean
logarithmic friction is fixed by scaling, while the dimensionless
factor $f$ depends only on the variance $s^2$ of the logarithmic
friction.

The function ${\cal D}(r,s^2)$ can be evaluated directly from
Eq.~\eqref{eq:logOU_Dcal} by averaging over stationary
Ornstein--Uhlenbeck trajectories. For the parameters used in
Fig.~\ref{fig:general_fast_error}(b), $m=1$, $\mu=0$, and $s^2=1$,
the midpoint condition gives
\begin{equation}
  f(1)=2.094,
\end{equation}
and hence
\begin{equation}
\tau_c=2.094.
\end{equation}
This value is used in the two-state crossover form shown by the solid
line in Fig.~\ref{fig:general_fast_error}(b), without any fitting
parameters.

\section{Numerical simulation details}
\label{app:numerical}

All simulations employed an analytic integration scheme based on the exact
solution of the linear Ornstein--Uhlenbeck Langevin equation. For the
dichotomous-friction model in Figs.~\ref{fig:relative_error} and
\ref{fig:Dunder_slow}, we exploited the fact that $\gamma$ is piecewise
constant within each interval. For the log-Ornstein--Uhlenbeck (log-OU)
friction model in Fig.~\ref{fig:general_fast_error}, the log-friction process
$Y(t)$ itself was updated exactly, while the velocity and position were
propagated using a quasi-exact scheme in which $\gamma$ was frozen at the
trapezoidal average of its values at the two endpoints of each substep.

The time discretization differed among the figures. For
Figs.~\ref{fig:relative_error}(a)--\ref{fig:relative_error}(c), rather
than using a fixed $\Delta t$, we used an event-driven scheme in which each
integration step was terminated exactly at the next switching time of the
dichotomous friction process. Because $\gamma$ does not change within such an
interval, the integration is exact and has no time-discretization error. For
Fig.~\ref{fig:Dunder_slow}, we instead used the piecewise-fixed adaptive
step size
\begin{equation}
  \Delta t=
  \begin{cases}
    10^{-3}, & t<1,\\
    10^{-2}, & 1\le t<10^2,\\
    10^{-1}, & 10^2\le t<10^4,\\
    1,       & t\ge10^4.
  \end{cases}
\end{equation}
For Fig.~\ref{fig:general_fast_error}, the substep size was
$\tau_{\rm env}/N_{\rm sub}$ with $N_{\rm sub}=20$. In
Fig.~\ref{fig:general_fast_error}(b), the smaller of $\tau_{\rm env}$ and the
overdamped relaxation time $\tau_{\rm rel}$ (the inverse of the overdamped
relaxation rate) was used in place of $\tau_{\rm env}$.

For Figs.~\ref{fig:relative_error}(a)--\ref{fig:relative_error}(c),
which show $\varepsilon_{\rm OD}/\varepsilon_{\rm OD}^{\rm fast}$ as a
function of $\tau_{\rm env}/\tau_c$ while scanning the particle mass, the
friction ratio, and $P_l$, respectively, each parameter point was averaged
over $N=40{,}000$ trajectories. The observation time was
\begin{equation}
  t_{\rm obs}=500\max\left(\frac{1}{\lambda_{\rm min}},\frac{1}{K}\right).
\end{equation}
Here $\lambda_{\rm min}=\min(|\lambda_+|,|\lambda_-|)$ is the slowest
velocity-correlation decay rate.
The dichotomous process was initialized according to its stationary
occupation probabilities, the velocity was drawn from the Maxwell--Boltzmann
equilibrium distribution, and the initial position was set to the origin.
Standard errors were estimated using a paired-difference estimator that
combined the overdamped and underdamped displacements. Their absolute values
were approximately $1\times10^{-3}$--$7\times10^{-3}$, corresponding to
roughly $0.1$--$1.5\%$ relative to $\varepsilon_{\rm OD}^{\rm sim}$, at every
point. The error bars are smaller than the symbols and are therefore not
visible in the figures, but they were retained in the analysis.

For Fig.~\ref{fig:Dunder_slow}, which shows the time evolution of
$D_{\rm eff}^{\rm under}(t,K)$ for three initial occupation probabilities,
we simulated $N=10{,}000$ trajectories per initial condition ($30{,}000$
trajectories in total). The velocity was initialized from the equilibrium
distribution and the position at the origin, with
$p_l(0)=0$, $P_l$, or $1$. Error bars were not computed for this figure. The
convergence of the three curves to within a few percent of one another at
long times suggests a statistical uncertainty of comparable magnitude.

For Fig.~\ref{fig:general_fast_error}(a), which shows
$\varepsilon_{\rm OD}^{\rm fast}$ as a function of $s^2$ at fixed
$\tau_{\rm env}=10^{-3}$, we used $N=10{,}000$ trajectories for each value
$s\in\{0.2,0.4,0.6,0.8,1.0,1.2,1.5,1.8,2.0\}$. For
Fig.~\ref{fig:general_fast_error}(b), which shows $\varepsilon_{\rm OD}$ as a
function of $\tau_{\rm env}$ at fixed $s=1$, we used $N=10{,}000$
trajectories at each of 19 logarithmically spaced values of
$\tau_{\rm env}\in[10^{-3},10^3]$. In both panels, $Y$ was initialized from
its stationary distribution $\mathcal{N}(\mu,s^2)$, the velocity from the
Maxwell--Boltzmann equilibrium distribution, and the position at the origin.
Error bars obtained by applying the delta method to a self-normalized ratio
estimator ranged from approximately $2.6\times10^{-3}$ to
$1.5\times10^{-2}$ in absolute terms.

\bibliography{akimoto}

\end{document}